\documentclass[journal]{IEEEtran}

\usepackage{cite}
\usepackage{amsmath,amssymb,amsfonts}
\usepackage{algorithmic}
\usepackage{graphicx}
\usepackage{textcomp}
\usepackage{xcolor}
\usepackage{url}
\usepackage[utf8]{inputenc}
\usepackage[normalem]{ulem}
\usepackage{booktabs}
\usepackage{colortbl} 
\usepackage{multirow} 
\usepackage{array}    
\usepackage{diagbox}
\newcommand\asif[1]{\textcolor{black}{#1}}
\begin{document}


\title{A Systematic Method for Optimum Biomedical Wireless Power Transfer using Inductive Links in Area-Constrained Implants}

\author{Asif~Iftekhar~Omi,~\IEEEmembership{Graduate Student Member,~IEEE}, Anyu~Jiang,~\IEEEmembership{Graduate Student Member,~IEEE,}~and~Baibhab~Chatterjee,~\IEEEmembership{Member,~IEEE}

\thanks{Manuscript received XXX XX, 202X; revised XXX XX, 202X; accepted XXX XX, 202X; Date of current version XXX XX, 202X. \textit(\textit{Corresponding author: Asif Iftekhar Omi})}

        
\thanks{The authors are with the Herbert Wertheim College of Engineering, Department of Electrical and Computer Engineering, University of Florida, Gainesville, FL 32611, USA. (e-mail: \{as.omi, chatterjee.b\}@ufl.edu)}

\thanks{Color versions of the figures are available at http://ieeexplore.ieee.org.}

\thanks{Digital Object Identifier 10.xxxx/TBCAS.202x.xxxxxxx}
 }
\maketitle


\begin{abstract}
In the context of implantable bioelectronics, this work provides new insights into maximizing biomedical wireless power transfer (BWPT) via the systematic development of inductive links. This approach addresses the specific challenges of power transfer efficiency (PTE) optimization within the spatial/area constraints of bio-implants embedded in tissue. Key contributions include the derivation of an optimal self-inductance with S-parameter-based analyses leading to the co-design of planar spiral coils and L-section impedance matching networks. To validate the proposed design methodology, two coil prototypes— one symmetric (type-1) and one asymmetric (type-2)— were fabricated and tested for PTE in pork tissue. Targeting a 20 MHz design frequency, the type-1 coil demonstrated a state-of-the-art PTE of $\sim$ 4\% (channel length = 15 mm) with a return loss (RL) $>$ 20 dB on both the input and output sides, within an area constraint of $<$  18 $ \times $  18  mm$^{2}$. In contrast, the type-2 coil achieved a PTE of $\sim$ 2\% with an RL $>$ 15 dB, for a smaller receiving coil area of $<$ 5x5 mm$^{2}$ for the same tissue environment. To complement the coils, we demonstrate a 65 nm test chip with an integrated energy harvester, which includes \asif{a} 30-stage rectifier and low-dropout regulator (LDO), producing a stable $\sim$ 1V DC output within tissue medium, matching theoretical predictions and simulations. Furthermore, we provide a robust and comprehensive guideline for advancing efficient inductive links for various BWPT applications, with shared resources in GitHub available for utilization by the broader community.

\end{abstract}

\vspace{-5mm}

\IEEEpeerreviewmaketitle

\section{Introduction}
\begin{figure}[t]
\centerline{\includegraphics[width=1\columnwidth]{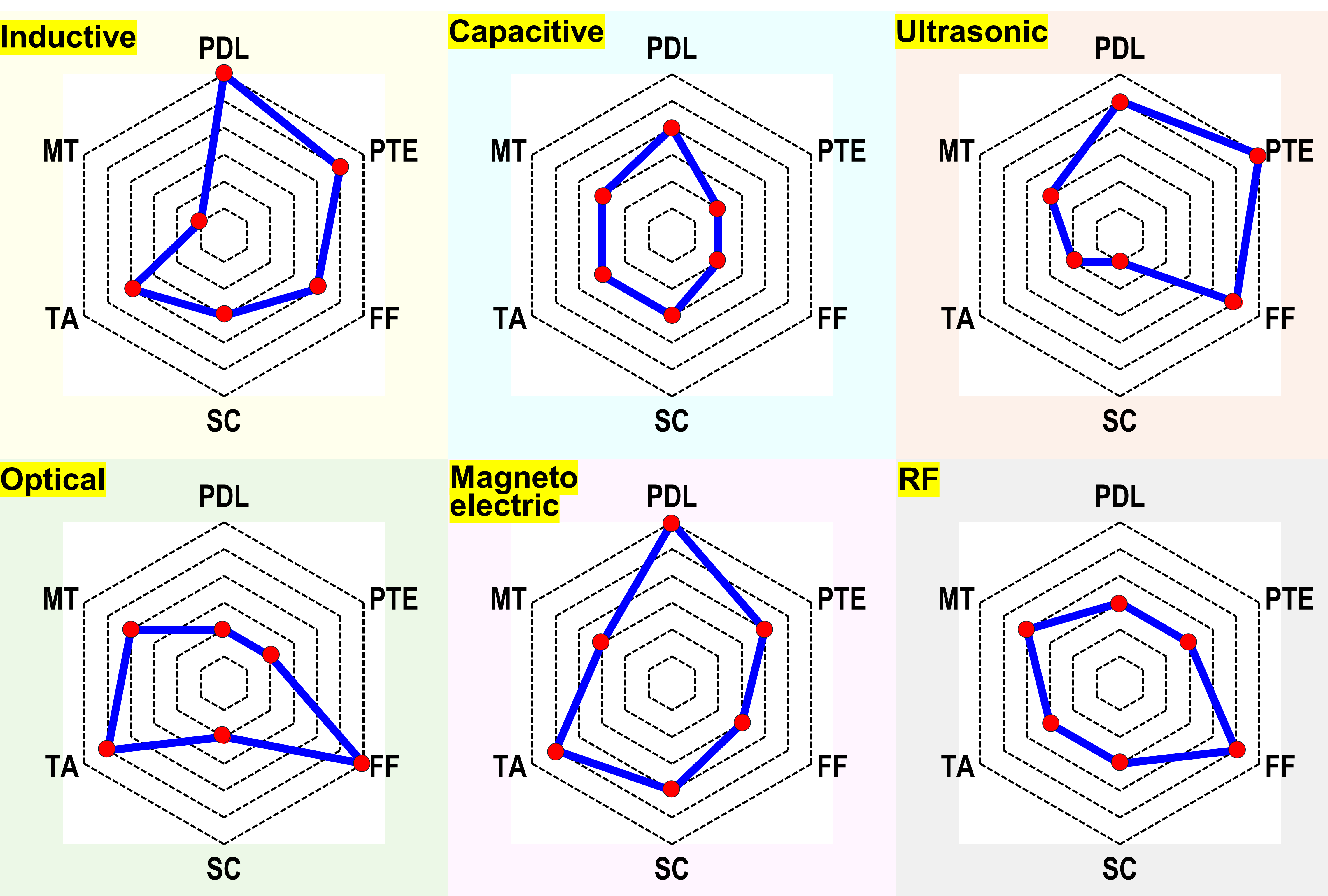}}
\vspace{-1mm}
\caption{Synopsis of various WPT modalities: brief comparison among the technologies in terms of power transfer efficiency, power delivered to the load, device form factor, surgical comfort, tissue absorption and misalignment tolerance where points away from the center indicate better performance (outermost hexagon represents the most desirable state). 
}
\label{intro1}
\vspace{-4mm}
\end{figure}
\begin{figure}[t]
\centerline{\includegraphics[width=1\columnwidth]{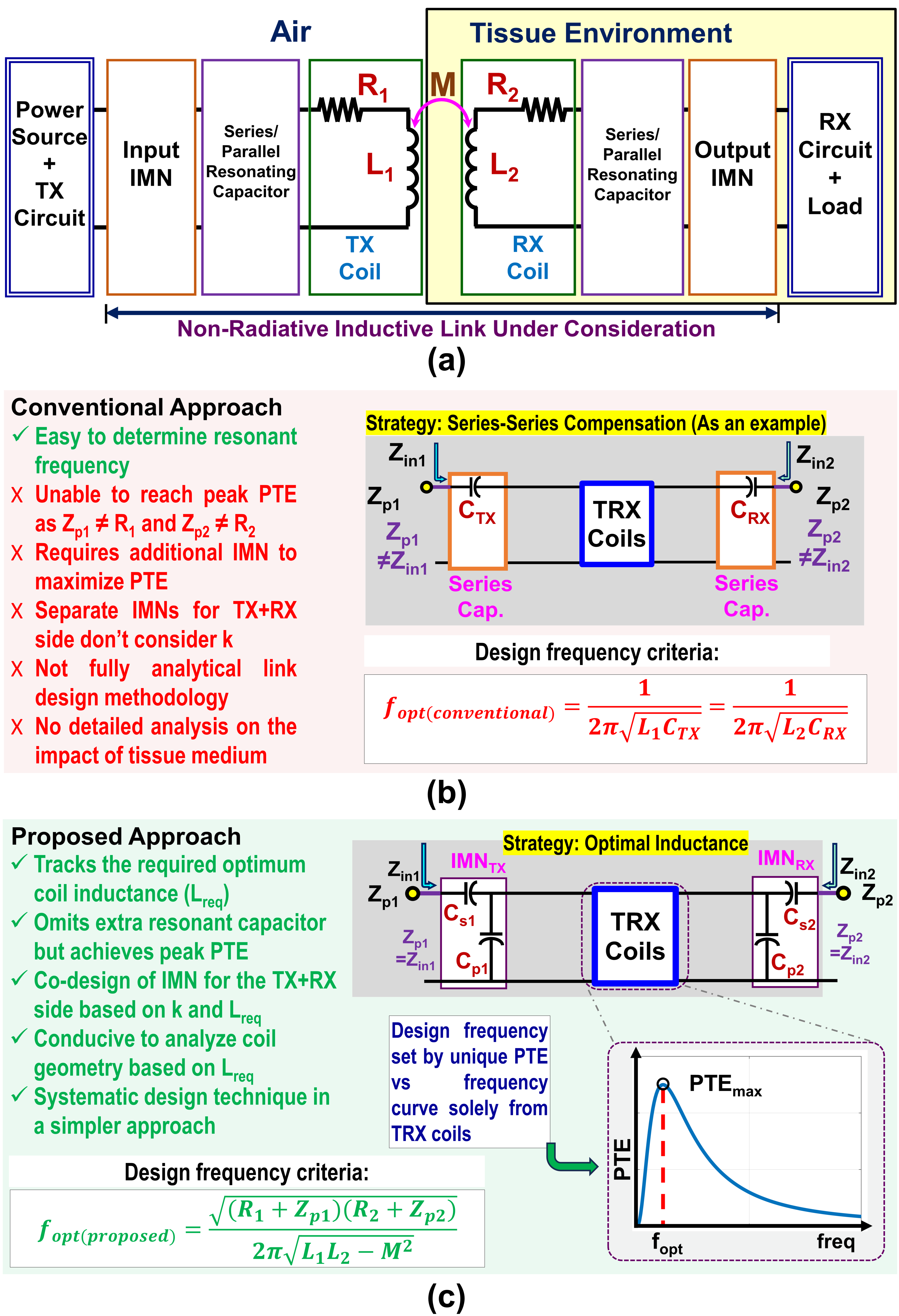}}
\vspace{-1mm}
\caption{\asif{Overview of NRIC link (widely adopted for biomedical implants): (a) generic block diagram, (b) conventional design approach, and (c) proposed design approach. The primary goal is to achieve optimal PTE of this NRIC link under conditions such as translation of circuit parameters into geometrical configurations, suitable IMN types, unequal port impedances, and spatial constraints imposed by the surrounding biological tissue environment.}}
\label{intro2}
\vspace{-3mm}
\end{figure}
\begin{figure*}[t]
\centerline{\includegraphics[width=1\textwidth]{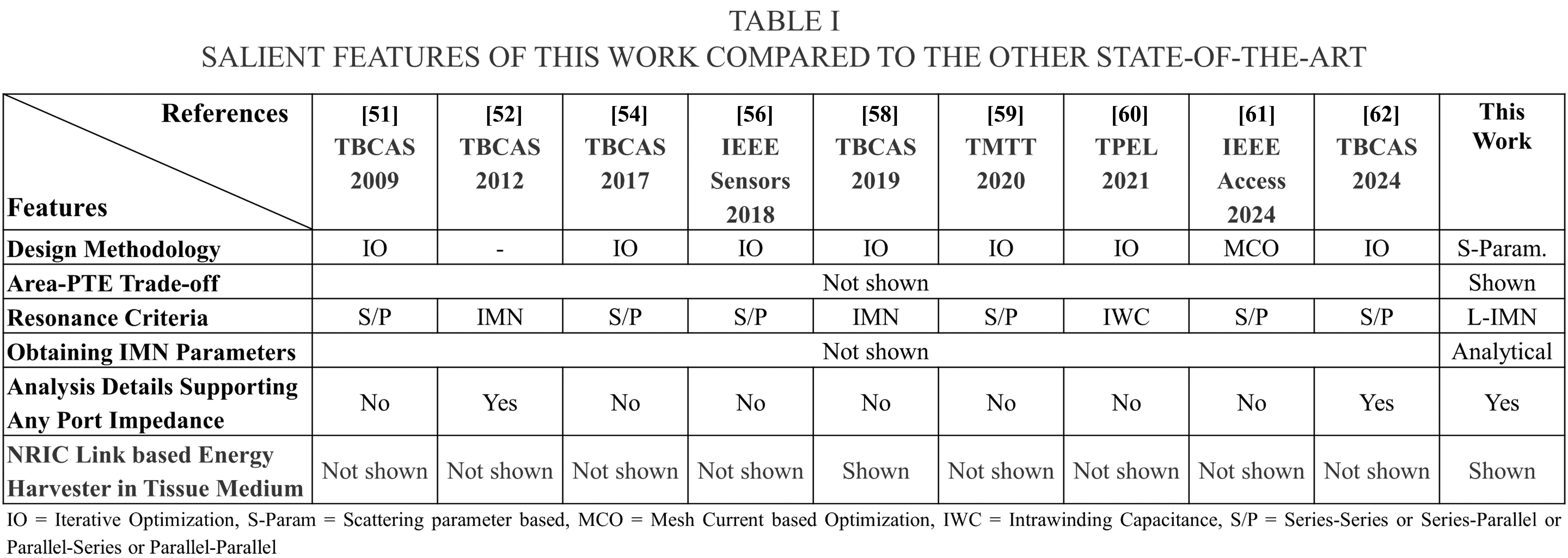}}
\vspace{-1mm}
\label{tab1}
\vspace{-1mm}
\end{figure*}

Advancements in bioelectronics have profoundly transformed contemporary implantable medical devices (IMDs), paving the way for enhanced management of chronic diseases and optimized surgical interventions \cite{chatterjee2023bioelectronic}. For example, neurostimulators \cite{barbruni2020miniaturised, chatterjee2023biphasic}, pacemakers \cite{monti2015resonant}, retinal implants \cite{chen2010integrated}, spinal cord stimulators \cite{chen2023intra}, and cochlear implants \cite{campi2021centralized} are some of the notable IMDs that have garnered considerable attention by providing targeted treatments. However, despite the splendid benefits offered to patients, powering these IMDs effectively continues to be one of the major challenges within the biomedical
circuits and system 
design community. 
Traditional percutaneous wiring methods provide a consistent power supply, but they are vulnerable to damage due to material incompatibility and scarring at the tissue interface, and maintaining sterility for these devices is an extra challenge \cite{haerinia2020wireless}. Similarly, internal power sources, like batteries, occupy substantial space and must be surgically replaced at regular intervals due to their limited charge capacity and recharge cycles \cite{hauser2021reliability}. Additionally, fluctuations in temperature and power consumption can lead to overheating, explosion, or chemical leakage. Recognizing these stringent limitations within the tissue environment, wireless power transfer (WPT) systems have emerged as key enablers for the progress of next-generation biomedical implants \cite{agarwal2017wireless,zhang2018wireless,monti2021wireless,roy2022powering}.
\subsection{Motivation and Significance}
Today's biomedical WPT (BWPT) for implantable applications encompass resonant inductive \cite{omi2024new, omi2024embc}, capacitive \cite{koruprolu2018capacitive}, ultrasonic \cite{ghanbari2019sub}, optical \cite{moon2021bridging}, magnetoelectric \cite{alrashdan2021wearable, hosur2023comparative}, and radio frequency based electromagnetic transmission \cite{das2017multiband} modalities. Accordingly, numerous studies have addressed the specific requirements of IMDs based on these diverse WPT systems characterized by power transfer efficiency (PTE), power delivered to the load (PDL), device form factor (FF), surgical comfort (SC), tissue absorption (TA) and misalignment tolerance (MT) \cite{singer2021wireless,chatterjee2023bioelectronic}. These parameters help to identify the pros and cons of the aforesaid untethered powering technologies as highlighted in Fig.~\ref{intro1}. Though another WPT method recognized as body-coupled powering (BCP) seems promising, it is still at the early exploration stage \cite{zhu2023biomedical, Modak_JSSC, chen2023intra, omi2024biocas}, contrary to body-coupled communication which has been explored in detail over the last decade by multiple groups including our own \cite{HBC1, HBC2, HBC3, HBC4, HBC5, HBC6, HBC7, HBC8, HBC9, HBC10, HBC11}. While multiple WPT options are prevalent, non-radiative inductive coupling (NRIC) based on resonant inductive WPT is considered the paragon among all as it is widely deployed in various IMD scenarios due to its size, scalable power range, superior short-distance power transmission performance, and tissue safety \cite{singer2021wireless}. A general block diagram of the NRIC link is depicted in  \asif{Fig.~\ref{intro2}(a)}, which is the prime focus of this work.
\subsection{Related Prior Works}
A review of the existing literature reveals a plethora of studies conducted thus far signifying extensive research on the design and implementation of versatile NRIC links \cite{zhu2015improving,villa2011high,zhao2017hybrid,park2011investigation,lim2014adaptive,beh2013automated,kim2015range,lee2016reconfigurable,lopez2000improvement,awuah2023novel,waters2014optimal,cove2016improving,kim2016free,kim2018asymmetric,5223583,zargham2012maximum,ibrahim2016figure,7926319,meng2016hybrid,ibrahim2018comprehensive,schormans2018practical,khalifa2019microbead,qiu2020digital,wang2020self,kobuchi2024automatic,saha2024multi,wu2024precise}. The conventional NRIC link is illustrated in Fig.~\ref{intro2}(b), which shows the resonating series/parallel (S/P) capacitors to the transmitter (TX) and receiver (RX) coils separately. The compensation configuration can be series-series (shown in Fig.~\ref{intro2}(b) as an example), series-parallel, parallel-series, and parallel-parallel. However, this resonant tank still produces suboptimal efficiency due to not fulfilling proper matching with the TX and RX circuitry unless additional IMNs are connected \cite{zargham2012maximum,khalifa2019microbead,pozar2011microwave}. This is due to the fact that the combined effect of non-zero TRX coil resistances and power transfer medium hinders the matching at both TX and RX ports, which has been elaborately described in Section II.E through an example. 

Beyond the traditional S/P compensation topologies, several advanced topologies have been introduced, including L-C-C \cite{zhu2015improving}, C-C-L \cite{villa2011high}, and hybrid configurations \cite{zhao2017hybrid}, which exhibit superior performance under misalignment conditions. The complexity of the circuit topology increases the system's susceptibility to parameter variations, thereby undermining its reliability and restricting any potential improvements in PTE. A comparable effect is observed when utilizing frequency tracking-based impedance matching networks (IMNs) \cite{park2011investigation,lim2014adaptive,beh2013automated} and coil repeaters \cite{kim2015range,lee2016reconfigurable} to facilitate impedance matching between the input impedance of the TX coil and the source impedance across a broad operational range. Even with the various IMNs on both the TX and RX sides, the entire design process for the transceiver (TRX) coils remains partially non-analytical. 

Meanwhile, numerous 
prior works have focused on improving PTE of WPT systems by optimizing the quality factor (Q) of TRX coils \cite{lopez2000improvement,awuah2023novel}, with techniques such as the divide-and-merge method proposed to reduce skin effects and enhance Q in thin-film printed coils, \asif{though} its application is limited to single-turn spiral designs. More recent approaches have explored optimizing coil geometries by varying width, pitch, and trace thickness \cite{waters2014optimal,cove2016improving,kim2016free}. \asif{These demonstrated significant resistance reduction and Q-factor improvement through trace width and turn-to-turn spacing adjustments.} Additionally, non-uniform coil turns and modifications to internal diameter to minimize losses, as well as the development of advanced design algorithms and geometric modifications \cite{kim2018asymmetric}, are observed to achieve optimal Q-factors. Nevertheless, none of these approaches comprehensively evaluate the overall system performance for BWPT. 

Conversely, some of the renowned studies \cite{5223583,ibrahim2016figure,7926319,meng2016hybrid,ibrahim2018comprehensive,schormans2018practical,khalifa2019microbead,qiu2020digital,wang2020self,kobuchi2024automatic,saha2024multi,wu2024precise,7015638,jiang2021efficient,zargham2014fully,biswas2019optimization,barbruni2024frequency,habibagahi2022design,del2023design} rely on iterative optimization-oriented design techniques to fully model the NRIC link with conventional S/P compensations, \asif{often assuming a fixed port impedance (typially $50~\Omega$).} While these models offer valuable insights, they may not reflect realistic tissue environments. Although \cite{zargham2012maximum} highlighted a closed-form analytical solution for the maximum achievable PTE in the entire link under arbitrary input impedance conditions, no guideline was provided \asif{on transforming} the circuit parameters into physical layout for optimum coil geometry. The most recent works such as \cite{wang2020self} \asif{propose} a novel planar coil design utilizing parallel-connected multilayer PCBs to generate significant parasitic capacitance for self-compensating resonance, while \cite{kobuchi2024automatic} applies mesh current distribution analysis to optimize TRX coil geometries, and \cite{saha2024multi} introduces a range-adaptive, flexible, human-specific, cluster-based NRIC link. Despite these accomplishments, they 
may not adequately address the specific criteria for coil parameters required to achieve resonant frequency from a coil-centric perspective and how the area constraints inside the tissue environment impact the PTE. Thus, 
a comprehensive and unified framework to optimize NRIC links with maximum design flexibility for implantable BWPT is still lacking as evident from this assessment of the previous related works.
\subsection{Our Contributions}
This paper introduces a unique systematic design methodology for any TRX coils, aiming to maximize PTE in biomedical implants under conditions where (1) the TRX coils are either symmetrical or asymmetrical, (2) the IMN type and parameters are unknown, (3) port impedances may be unequal, and (4) the constraints of the biological tissue environment limit the coil area. The proposed design concept is highlighted in \asif{Fig.~\ref{intro2}(c)}, and the key contributions compared to the other state-of-the-art in solving the specific research problem as outlined earlier are further summarized in Table I. 
Realizing that the efficiency for any NRIC link peaks at a particular frequency, even without considering any resonant capacitors/IMN, we co-design the coil and the IMN at that target frequency.
The underlying idea is to deduce that exclusive frequency, dependent on the inherent coil parameters, and then apply the suitable IMN to achieve the highest PTE by ensuring the minimum return loss between the TX and RX. A detailed theoretical analysis is carried out to obtain the closed-form design equations in terms of S-parameters to determine the design parameters of the NRIC link. 
Specifically, the contributions of this work is multi-fold:

\begin{itemize}
\item
\textit{Deriving the optimum self-inductance ($L_{opt}$)} as a function of the inherent coil parameters and the area available, based on the target frequency for WPT.
\end{itemize}
\begin{itemize}
\item
\textit{Devising both symmetric and asymmetric TRX coils utilizing the identified $L_{opt}$}, while inherently co-designing matching network requirements.
\end{itemize}
\begin{itemize}
\item
\textit{Developing a systematic approach to design the IMN for arbitrary port impedance} based on the derived $L_{opt}$.
\end{itemize}
\begin{itemize}
\item
\textit{Analyzing the trade-off between coil area and PTE} for different implantable applications.
\end{itemize}
\begin{itemize}
\item
\textit{The fully automated design approach is shared on GitHub} \cite{Sym-Asym-TRX-Coils} as an open-source tool for utilization by the greater biomedical system design community.
\end{itemize}

Although the proposed methodology adopting these features in an NRIC link design is generic, planar spiral-type coils are chosen for demonstration purposes due to their suitability for IMDs. While centered on biological and healthcare contexts, the design principles hold broader relevance for other WPT systems. In contrast to our prior works \cite{omi2024new,omi2024embc}, this paper introduces the following advancements:
(a) addressing the prevailing critical challenges in the existing state-of-the-art biomedical NRIC link design,
(b) extending the theoretical analysis to formulate a more generalized design solution, (c) refining design methodology for both symmetric and asymmetric coils with all possible L-type IMNs, (d) explicating coil area dependence in terms of S-parameters, (e) co-designing of an energy harvester system with TRX coils, and (f) providing detailed measurement results obtained from the coil PCBs integrated with the energy harvester system (in the form of a 65 nm test-chip) validating the correlation among theory, simulation as well as measurements.  
\subsection{Organization of Paper}
The remainder of this paper is organized as follows.
In Section II, the theory and design equations of the proposed NRIC link are developed, and subsequently, the derivation of the corresponding S-parameters, the area-PTE trade-off, and the IMN requirements are presented with the integrated energy harvester.  Then, the detailed step-by-step design
procedure of the complete NRIC link inside tissue environment is explained in Section III, and a few examples are demonstrated to support the theoretical design capability. Afterward, the fabricated prototypes and their EM simulated, as well as the measured results, are demonstrated in Section IV. The performance of the energy harvester chip is also presented. Finally, the overall achievement of this work is compared with some of the state-of-the-art techniques, followed by our conclusions in Section V.

\section{Proposed Analytical Design Equations}
Starting with the standard two-coil NRIC system, we explore coil design constraints and provide a simplified analysis of the IMN, all aimed at achieving an optimized PTE condition within the tissue environment.
\vspace{-3mm}
\subsection{Finding the Optimum Frequency}
\begin{figure}[t]
\centerline{\includegraphics[width=0.9\columnwidth]{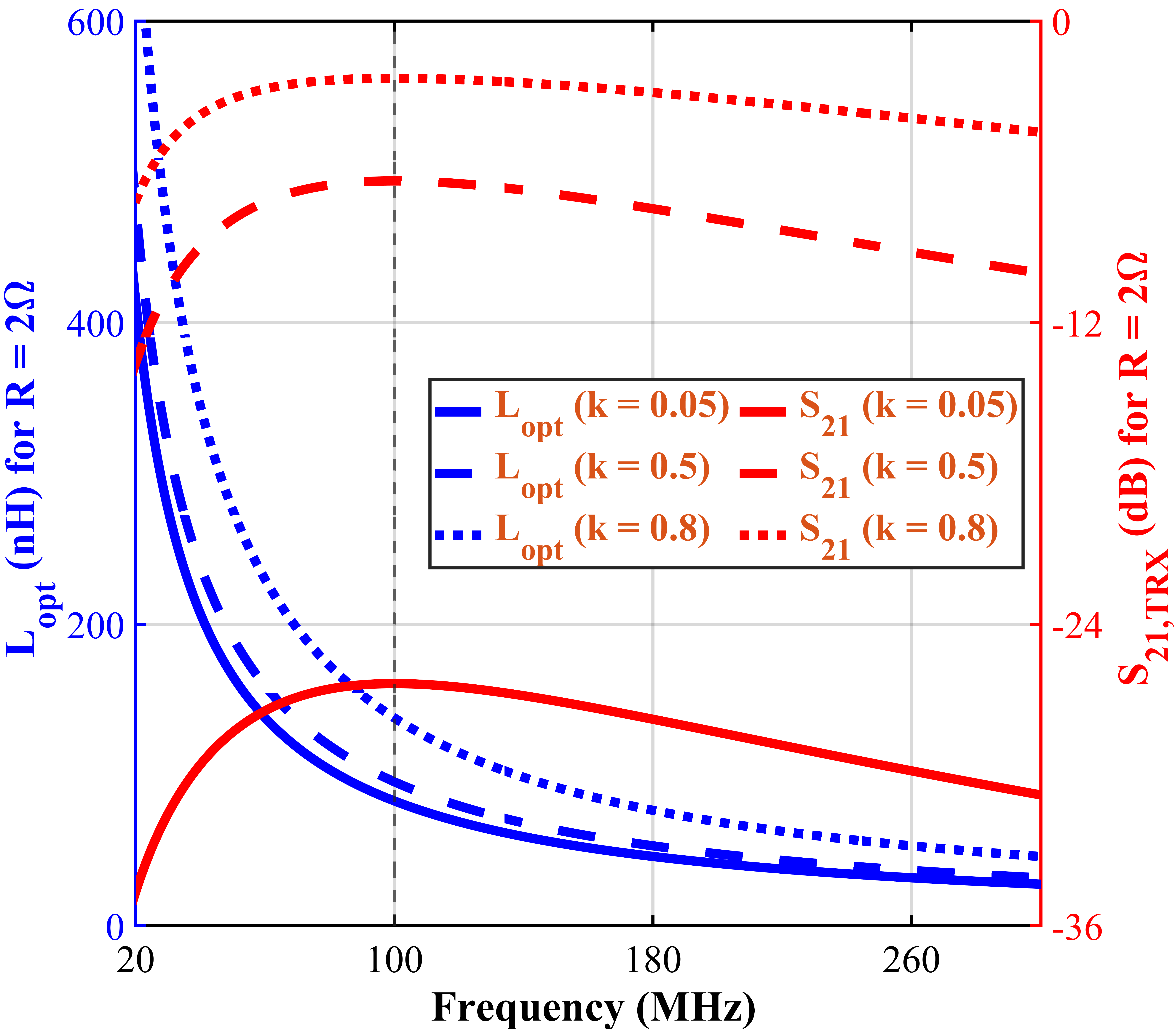}}
\vspace{-3mm}
\caption{Estimate of optimum inductance for peak PTE: $L_{opt}$ is tracked at each frequency to find the best possible inductance for each frequency and initiate the design process for the targeted coupling co-efficient indicative of the TX-RX distance.}
\label{2.1.1}
\end{figure}
\begin{figure}[t]
\centerline{\includegraphics[width=1\columnwidth]{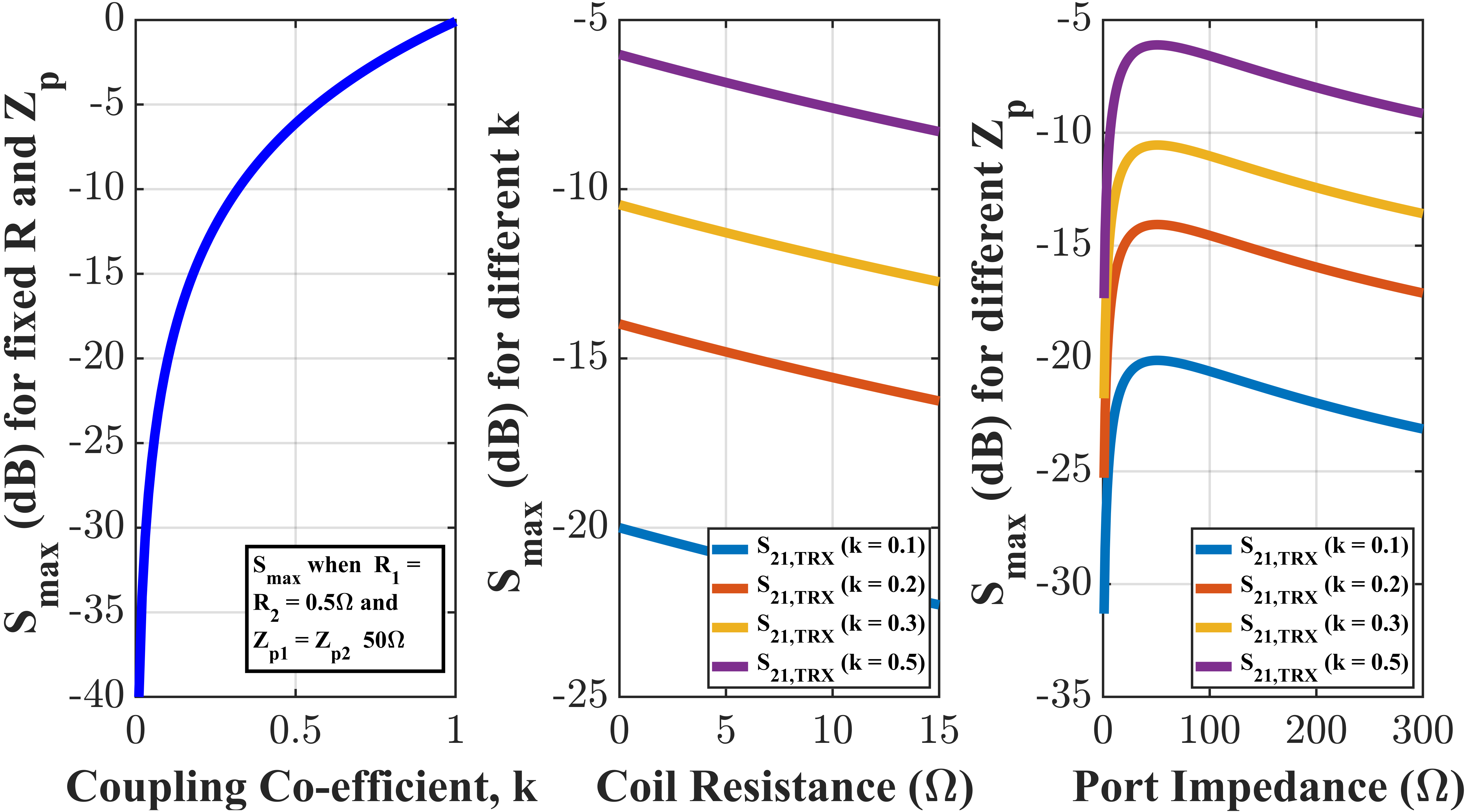}}
\vspace{-2mm}
\caption{Variation of $\left | S_{max} \right |$ (= Maximum $\left | S_{21,TRX} \right |$) in terms of (a) coupling coefficient, $k$  (b) coil resistance, $R_{j}$, and (c) unequal port impedances, $Z_{p}$. This provides a preliminary idea about the peak performance achievable from the TRX coils }
\label{2.1.2}
\vspace{-4mm}
\end{figure}

For the inductively coupled coils separated by a distance $d$, the characteristic Z-matrix can be defined as follows \cite{zhang2018wireless}.
\begin{equation}
Z_{coil}=\begin{bmatrix}
R_{1}+j\omega L_{1} & j\omega k\sqrt{L_{1}L_{2}}
 \\ 
j\omega k\sqrt{L_{1}L_{2}}
 & R_{2}+j\omega L_{2}
\end{bmatrix}\label{eq}
\end{equation}

Here, $L_{j}$ represents the self-inductances, while $R_{j}$ characterizes coil resistance and $j$ $\in$ $\left \{ 1,2 \right \}$ delineating the TX and RX coils, respectively. The coupling coefficient, $k$ between the coils is defined in terms of mutual inductance, $M$ where $M=k\sqrt{L_{1}L_{2}}$. Subsequently, transforming (1) into the S-parameter matrix yields $\left | S_{21,TRX} \right |$, indicative of power transfer efficiency (PTE) for the TRX coils.
\begin{equation}
\left | S_{21,TRX} \right |=\frac{\omega t_{1}}{\sqrt{(t_{2}-t_{3}\omega ^{2})^{2}+\omega ^{2}t_{4}^{2}}}\label{eq}
\end{equation}

where,
\begin{equation}
t_{1}=2M\sqrt{Z_{p_{1}}Z_{p_{2}}}
\label{eq}
\end{equation}
\begin{equation}
t_{2}=(R_{1}+Z_{p_{1}})(R_{2}+Z_{p_{2}})\label{eq}
\end{equation}
\begin{equation}
t_{3}=L_{1}L_{2}-M^{2}\label{eq}
\end{equation}
\begin{equation}
t_{4}=L_{1}(R_{2}+Z_{p_{2}})+L_{2}(R_{1}+Z_{p_{1}})\label{eq}
\end{equation}

Within this framework, $Z_{p_{1}}$ and $Z_{p_{2}}$ signify port impedances at the TX and RX ends, respectively. A detailed examination of these parameters highlights an intrinsic frequency, $f_{opt}$, at which $\left | S_{21} \right |$ achieves its inherent peak, $S _{max}$, as also depicted in Fig.~\ref{2.1.1}.

Now setting, $\frac{\mathrm{d} }{\mathrm{d} f}(\left | S_{21,TRX} \right |)=0$, leads to the derivation of$f_{opt}$ as outlined in (7). The details are thoroughly outlined in Appendix A.
\begin{equation}
f_{opt}=\frac{1}{2\pi }\sqrt{\frac{t_{2}}{t_{3}}}\label{eq}
\end{equation}

At this $f_{opt}$, maximum $\left | S_{21,TRX} \right |$ can be denoted as $\left | S_{max} \right |$, which is derived in (8).
\begin{equation}
S_{max}=\frac{t_{1}}{t_{4} }\label{eq}
\end{equation}

Also, Fig.~\ref{2.1.2} indicates the best possible PTE from TRX coils as a function of $k$, $R_{j}$ and $Z_{p}$. This is visible that $\left | S_{max} \right |$ increases from -40 dB to 0 dB for 0 $<$ $k$ $<$ 1 where $R_{j}$ and $Z_{pj}$ are kept constant at $0.5~\Omega$ and $50~\Omega$ respectively. While varying $R_{j}$ ($R_{1}$ = $R_{2}$), $\left | S_{max} \right |$ is reduced when $R_{j}$ becomes higher, which is expected. Also, when the shifting of $\left | S_{max} \right |$ is evident for unequal port impedances when $Z_{p1}$ is kept fixed at $50~\Omega$ and the $Z_{p2}$ is varied.

Further exploration into (7) presents $L_{opt}$ for an NRIC system shown in (9).
\begin{equation}
L_{opt}=\sqrt{\frac{(R_{1}+Z_{p_{1}})(R_{2}+Z_{p_{2}})}{4\pi^{2} f_{opt}^{2}(1-k^{2})}}\label{eq}
\end{equation}

This leads to two possible cases.

\textit{Case-1: Symmetric TRX Coils ($L_{1}$ = $L_{2}$)}

This indicates TRX coils of the same sizes where geometric shapes at both ends can be different as long as their area is equal. Thus, the values of $L_{1}$ and $L_{2}$ can be calculated from (9) as follows:
\begin{equation}
L_{opt}={L_{1}=L_{2}}\label{eq}
\end{equation}

\textit{Case-2: Asymmetric TRX Coils ($L_{1}$ $\neq~$$L_{2}$)}

This indicates TRX coils with unequal area (preferably, TX coil has a much higher area than RX coil for biomedical implants). Thus, it gives an additional degree of freedom as one of the inductance values can be chosen, and the other can be calculated from (9) as shown in (11).
\begin{equation}
L_{opt}=\sqrt{L_{1}L_{2}}\label{eq}
\end{equation}

Deriving $f_{opt}$ and $L_{opt}$ concurrently from S-parameters, as observed here, establishes a theoretical cornerstone for understanding the NRIC link and initiates the coil design process, setting it apart from conventional methods.
\vspace{-3mm}
\subsection{Area-PTE Trade-off}
\begin{figure}[t]
\centerline{\includegraphics[width=1\columnwidth]{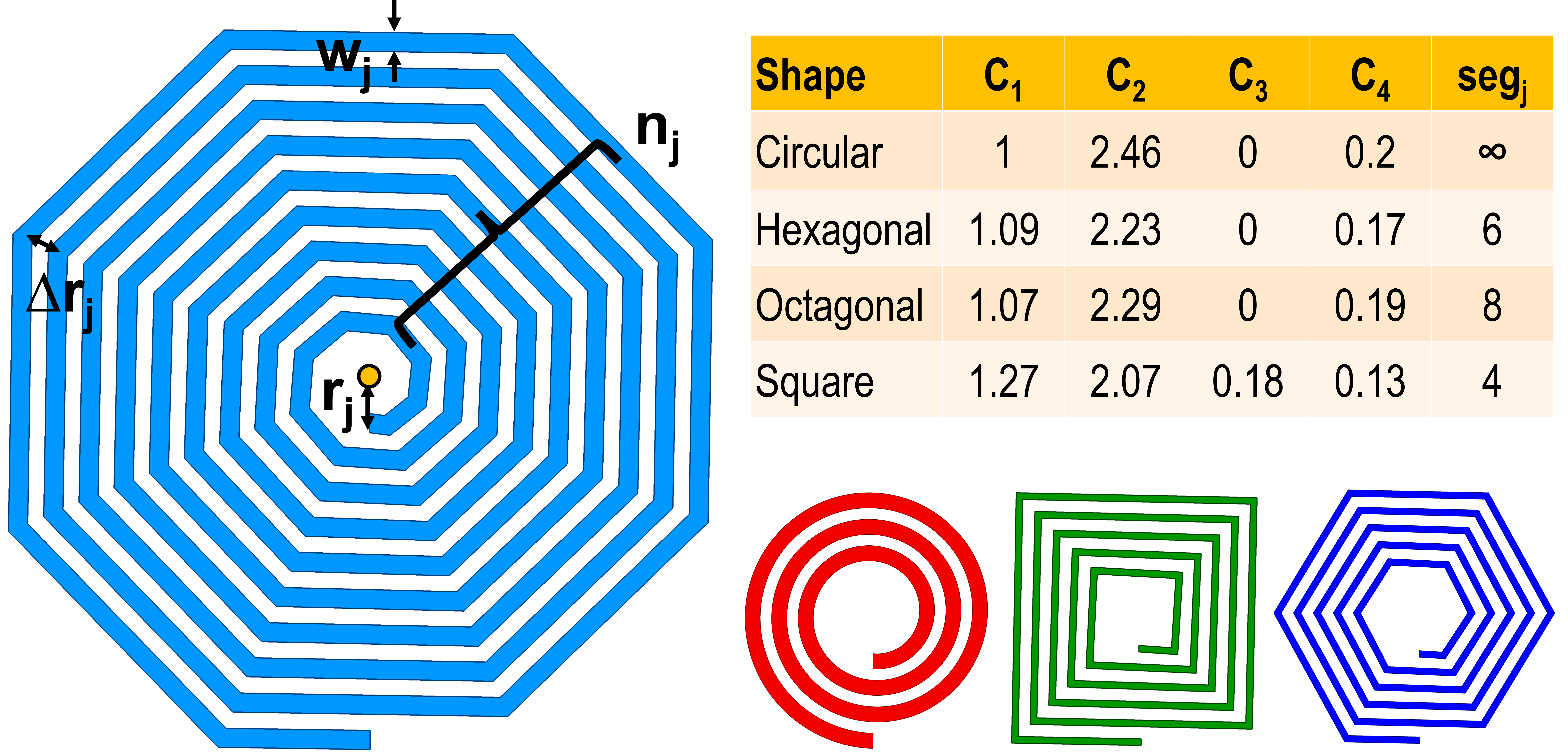}}
\caption{Arbitrary coil structures showing the physical parameters achieved through the modified Wheeler expression or the current sheet approximation method \cite{mohan1999simple}. Given the enhanced accuracy of the latter method, our analysis predominantly employs this approach for determining the inductance of planar spiral coils.}
\label{theory2}
\end{figure}

\begin{figure}[htbp]
\centerline{\includegraphics[width=1\columnwidth]{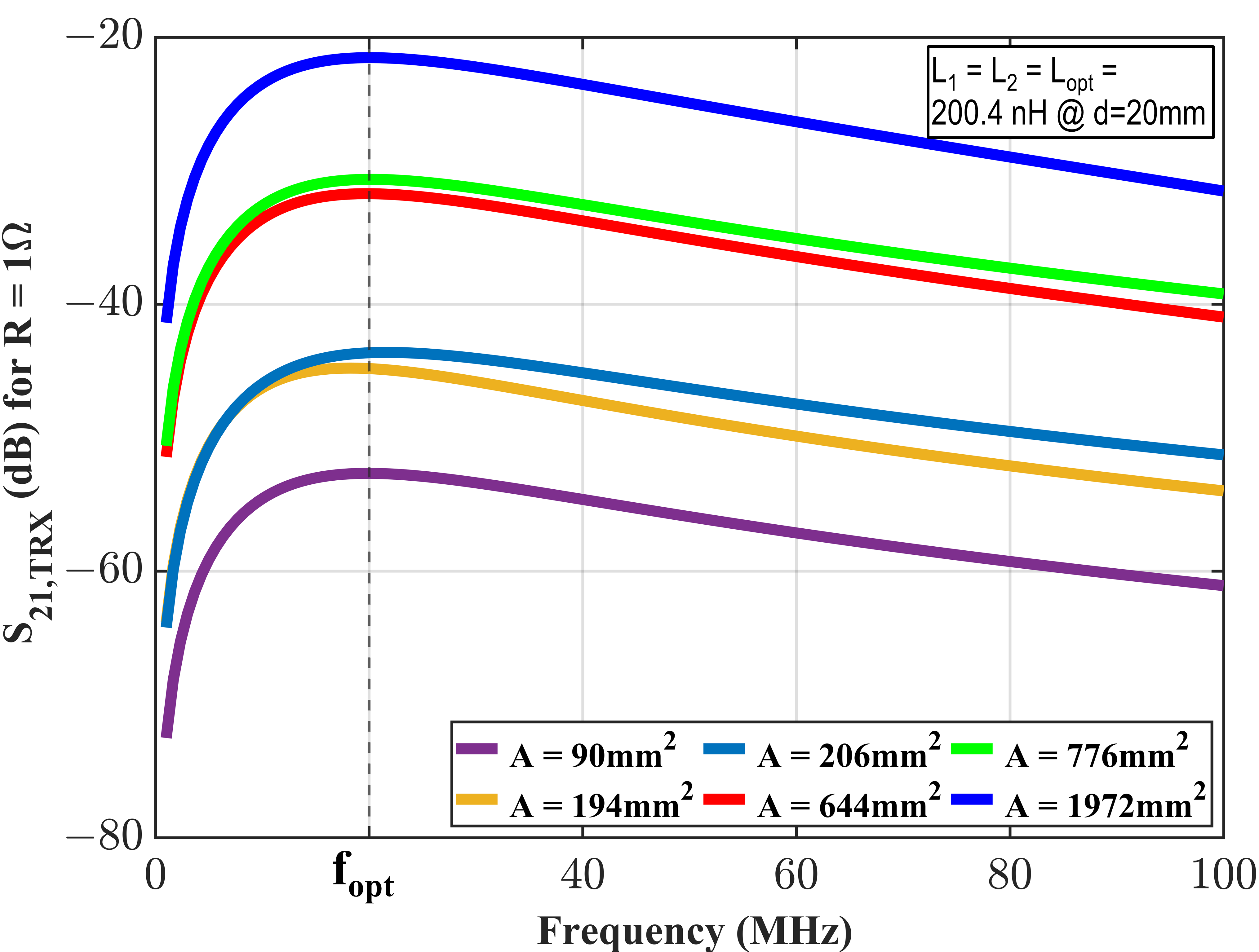}}
\vspace{-2mm}
\caption{Impact of coil area on PTE: maximizing $\left | S_{21} \right |$ necessitates a larger coil area. It stems from the fact that increased coil areas, despite having similar inductance values, yield enhanced coupling coefficients for a consistent $L_{opt}$.}
\label{theory3}
\vspace{-2mm}
\end{figure}

\begin{figure}[htbp]
\centerline{\includegraphics[width=0.9\columnwidth]{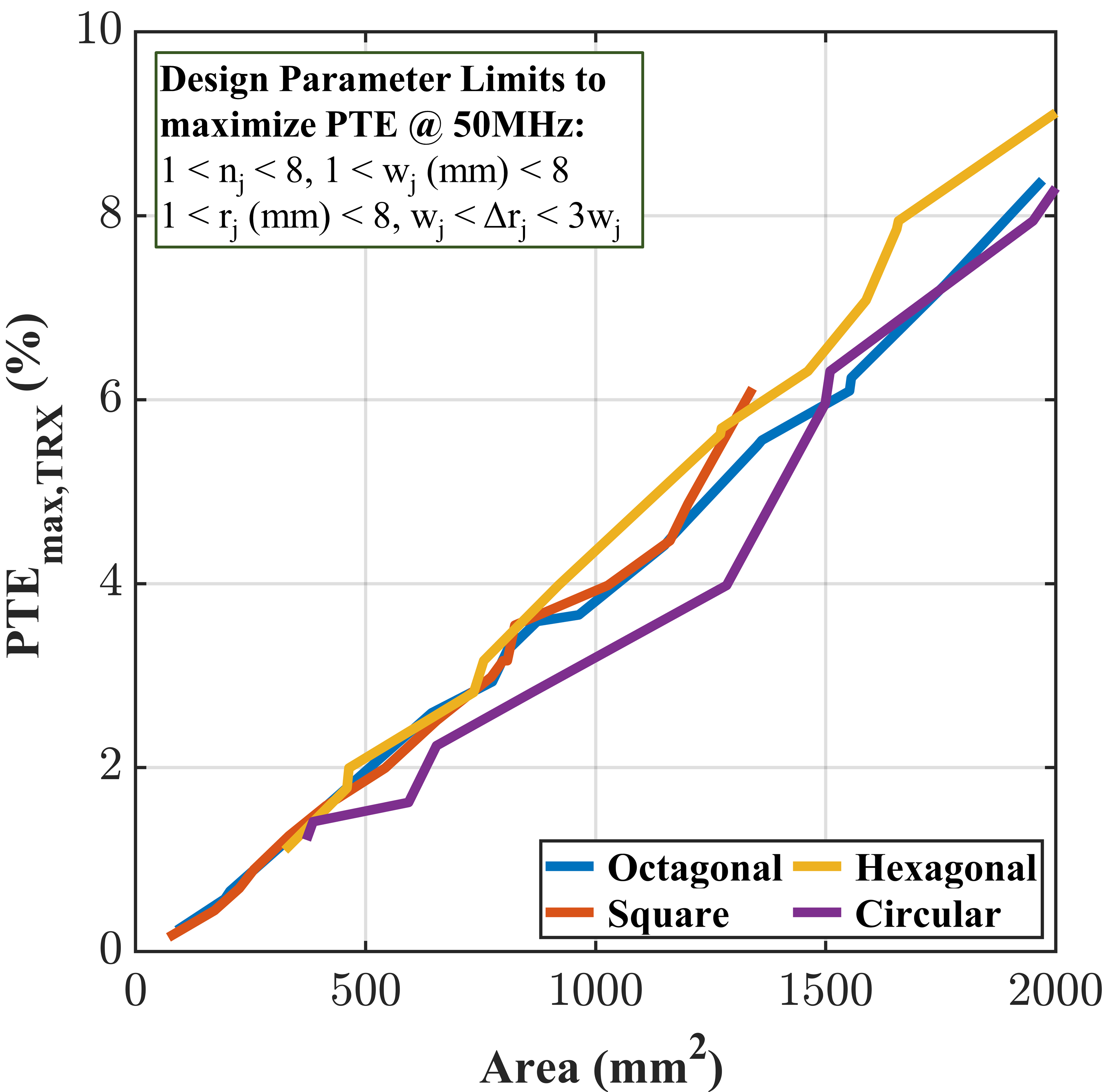}}
\vspace{-2mm}
\caption{Estimate of peak PTE for variable shaped geometry: $PTE_{max,TRX}$ (= maximum PTE of TRX coils) is obtained under the same design constraints. Altering the shape doesn't add any significant value in terms of PTE obtained according to the PTE formula described in Appendix C. Instead of circuit parameters, this comparative $PTE_{max,TRX}$ in terms of physical parameters provides further insight into the limit of area for any specific implants.}
\label{theory4}
\vspace{-2mm}
\end{figure} 

Identifying the perfect coil geometry is another ubiquitous challenge 
for BWPT. As discussed in Section I, most of the reported works are iterative optimization-based. But with the proposed $L_{opt}$ in our case, the next step is to translate it into approximate physical geometry. For the planar spiral coil, $L_{opt}$ can be further formulated as follows \cite{mohan1999simple}.
\begin{equation}
L_{opt}=\frac{C_{1}\mu _{0}\mu _{r}n_{j}^{2}d_{avg_{j}}}{2}[\ln (\frac{C_{2}}{\phi_{j} })+C_{3}\phi_{j} +C_{4}\phi_{j}^{2}]\label{eq}
\end{equation}

where, $d_{avg_{j}}$ denotes the average diameter, and $\phi_{j} $ signifies the filling ratio. The coefficients $C_{1}$, $C_{2}$, $C_{3}$, and $C_{4}$ are defined for geometric shapes such as square, hexagonal, octagonal, and circular planar spiral coilss, as mentioned in \cite{mohan1999simple}. Yet, these coefficients can be ascertained for arbitrary shapes utilizing finite element method (FEM) tools, notably Ansys HFSS. Given $r_{j}$ as the initial radius, $seg_{j}$ indicating the polygon's order, $n_{j}$ as the turn count, $\Delta r_{j}$ as the turn-by-turn radius increment, and $w_{j}$ as the coil width per turn, we infer the values of $d_{avg}$, $\phi_{j} $, and area, $A_{j}$, as demonstrated in (13) to (15). The coil parameters are properly represented in Fig.~\ref{theory2}.
\begin{equation}
\varphi_{j} =\left ( \frac{\sqrt{A_{j}}}{d_{avg_{j}}}-1 \right )\label{eq}
\end{equation}
\begin{equation}
d_{avg_{j}}=(2r_{j}+n_{j}\Delta r_{j})\cos (\frac{\pi }{seg_{j}})\label{eq}
\end{equation}
\begin{equation}
A_{j} =[w_{j}+2(r_{j}+n_{j}\Delta r_{j})\cos (\frac{\pi }{seg_{j}}))]^{2}\label{eq}
\end{equation}

An intriguing observation from equations (13) to (14) is the potential for different parameter values to produce the same $L_{opt}$ where coil area directly impacts PTE. For example, when the TRX coils are placed at 20mm apart with diverse sizes, producing $L_{opt}$ = 200.4nH, a consequential trade-off is observed in Fig.~\ref{theory3}. Further, it is also evident from Fig.~\ref{theory4} that modifying the shape doesn't necessarily indicate better PTE in terms of area varied for different shapes to maximize PTE at 50 MHz (as another design example). If we aim for higher PTE, the area will also increase. However, it can be inferred from this analysis how much PTE is obtainable within a range of coil sizing. This interplay emphasizes careful consideration before IMN implementation.
\vspace{-3mm}

{\subsection{Mapping Coil Geometry to Circuit Parameters}
After deciding the TRX coil geometries with area-PTE trade-off consideration, this sub-section outlines obtaining and validating the desired value of $L_{j}$, $R_{j}$, and $k$. Now, the ABCD matrix for the TRX coils can be found by translating (1) into the corresponding ABCD parameters as shown in (16)  \cite{pozar2011microwave}.
\begin{equation}
[T_{TRX}] =\begin{bmatrix}
A_{coil} & B_{coil} \\ 
C_{coil} & D_{coil}
\end{bmatrix}\nonumber \\
\end{equation}
\begin{equation}
=
\begin{bmatrix}
\frac{R_{1}+j\omega L_{1}}{j\omega M} & \frac{\omega ^{2}M^{2}+(R_{1}+j\omega L_{1})(R_{2}+j\omega L_{2})}{j\omega M}
 \\ 
\frac{1}{j\omega M}
 & \frac{R_{2}+j\omega L_{2}}{j\omega M}
\end{bmatrix} \label{eq}
\end{equation}

Now, the electrical parameters ($L_{j}, R_{j}$ and k) of the TRX coils can be obtained from these ABCD parameters of the coils as deduced in (17)-(20). 
\begin{equation}
L_{1}=\frac{Im\left\{Z_{p1} (\frac{pv+x}{qv-x})\right\}}{2\pi f}\label{eq}
\end{equation}
\begin{equation}
R_{1}=Re\left\{ (\frac{pv+x}{qv-x})\right\}\label{eq}
\end{equation}
\begin{equation}
L_{2}=\frac{Im\left\{Z_{p2} (\frac{qu+x}{qv-x})\right\}}{2\pi f}\label{eq}
\end{equation}
\begin{equation}
R_{2}=Re\left\{ (\frac{qu+x}{qv-x})\right\}\label{eq}
\end{equation}
\begin{equation}
k=\left [ Re\left\{ A_{coil}\right\}Re\left\{ D_{coil}\right\} \right ]^{-\frac{1}{2}}\label{eq} 
\end{equation}

where,
\begin{equation}
p=1+S_{11,TRX}\label{eq}
\end{equation}
\begin{equation}
q=1-S_{11,TRX}\label{eq}
\end{equation}
\begin{equation}
u=1+S_{22,TRX}\label{eq}
\end{equation}
\begin{equation}
v=1-S_{22,TRX}\label{eq}
\end{equation}
\begin{equation}
x=S_{12,TRX}S_{21,TRX}\label{eq}
\end{equation}

A detailed derivation of (17)-(21) is available in Appendix B. Thus, these obtained $L_{j}$, $R_{j}$, and $k$ further ensure that the optimal inductance is accurately modeled into the geometrical parameters of the spiral coil structure. Simultaneously, the required conversion of the remaining electrical parameters into the equivalent layout is confirmed.
\vspace{2mm}
\subsection{Impact of Tissue Medium}
\begin{figure}[t]
\centerline{\includegraphics[width=1\columnwidth]{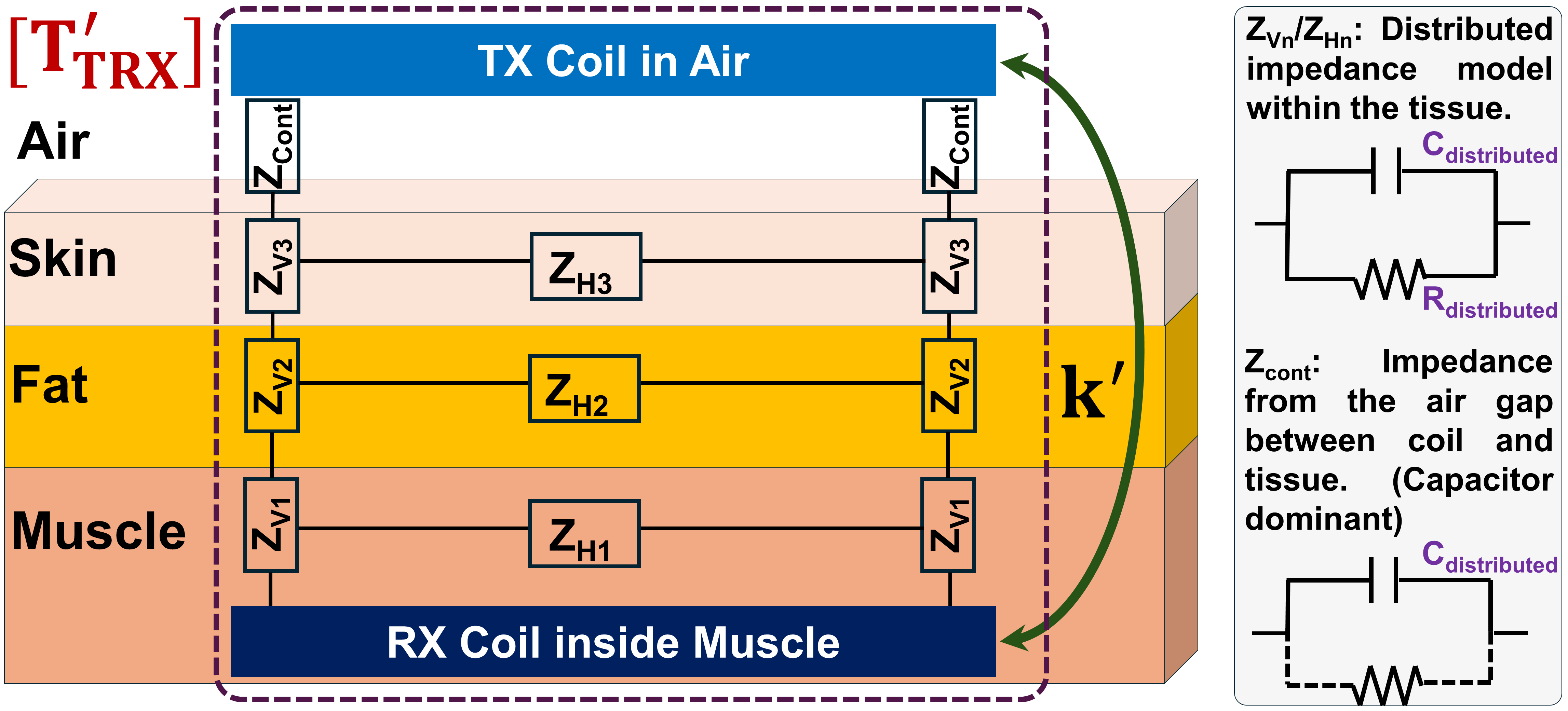}}
\vspace{-1mm}
\caption{Impact of heterogeneous tissue channel: tissue's magnetic properties do not alter magnetic fields, yet its conductivity can induce eddy currents that counteract fields from the TX coil, reducing the field at the RX coil. This effect, depicted as $Z_{Vn}$ and $Z_{Hn}$, necessitates adjusting $T_{TRX}$ in (16) to retain the analytical design formulations.}
\label{trmed}
\vspace{-2mm}
\end{figure}
\begin{figure}[t]
\centerline{\includegraphics[width=1\columnwidth]{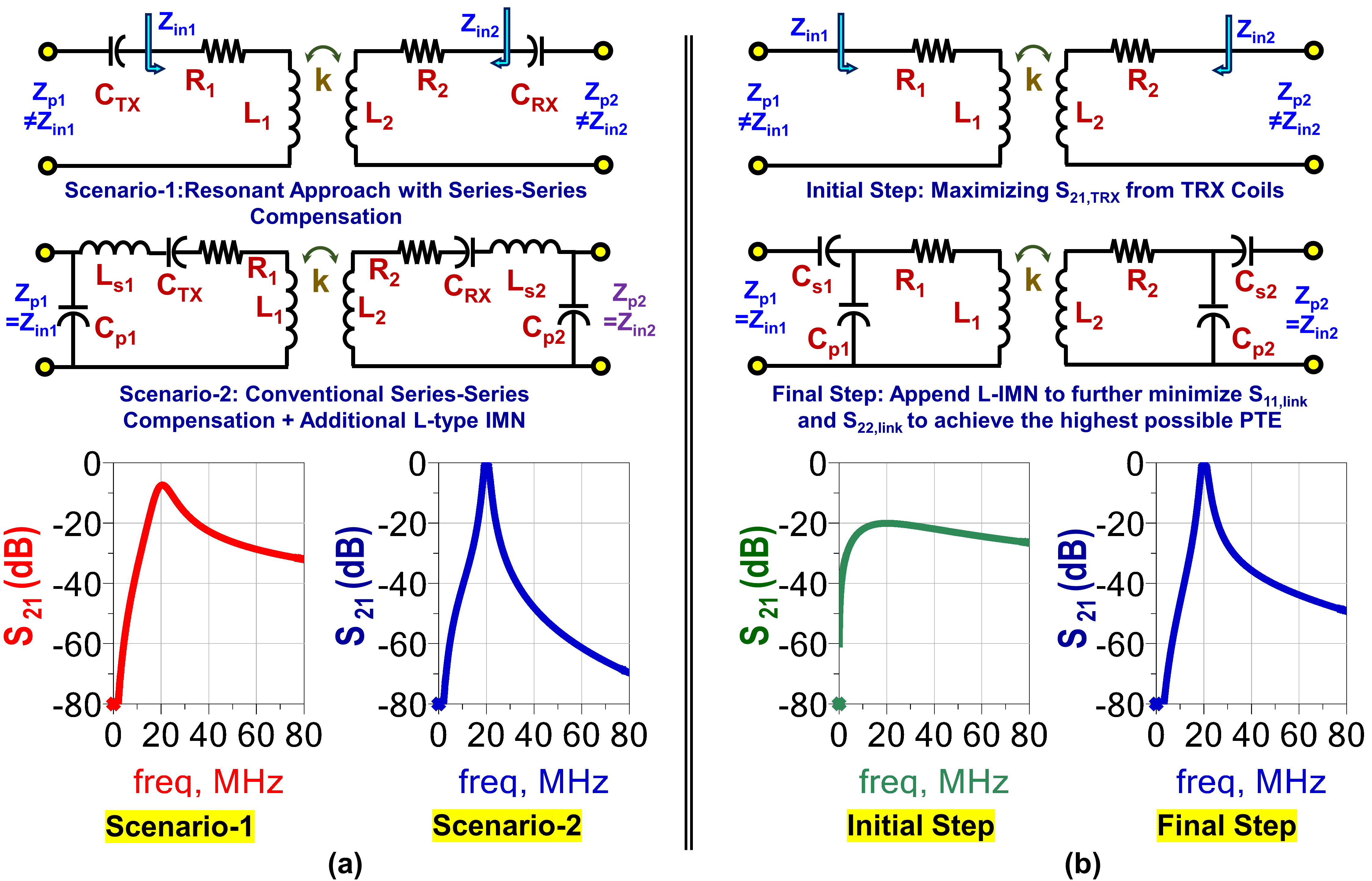}}
\vspace{-1mm}
\caption{IMN vs Series/Parallel compensation: (a) the left side indicates two scenarios, one with resonant approach as series-series compensation and another with additional L-type IMN, and (b) the right side indicates the proposed approach.}
\label{LIMN_0}
\end{figure}
\begin{figure}[htbp]
\centerline{\includegraphics[width=1\columnwidth]{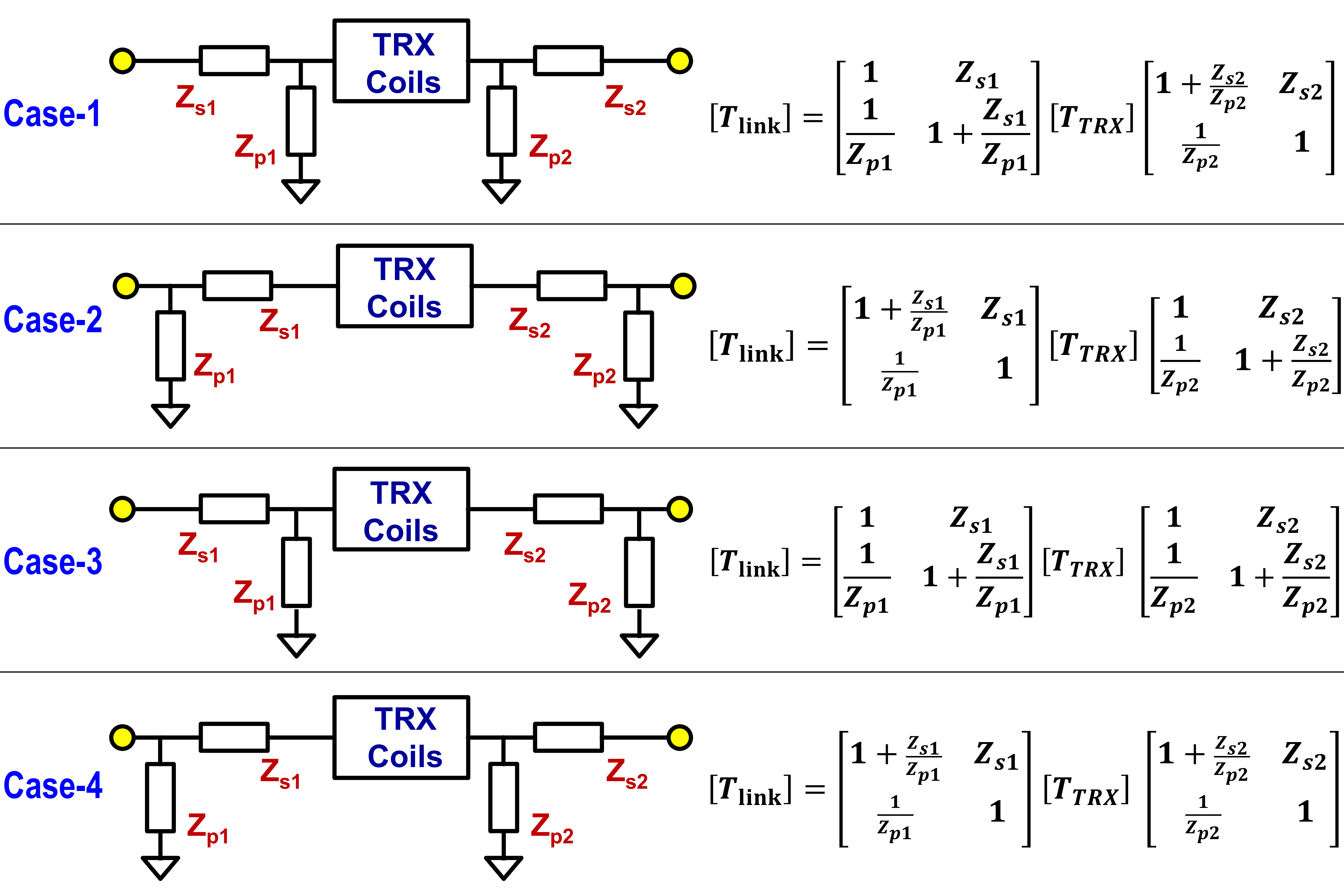}}
\vspace{-1mm}
\caption{Impact of different L-type IMN configurations: transmission matrix for the entire NRIC link, [$T_{link}$] is formulated for four distinct cases, which lead to $4 \times 2^{4} = 64$ possible combinations consisting of capacitors or inductors. Depending on the TRX coil geometry and distributed tissue impedance, a suitable IMN can be established out of these possible options.}
\label{LIMN_1}
\vspace{-2mm}
\end{figure}
\begin{figure*}[htbp]
\centering
\includegraphics[width=0.8\textwidth]{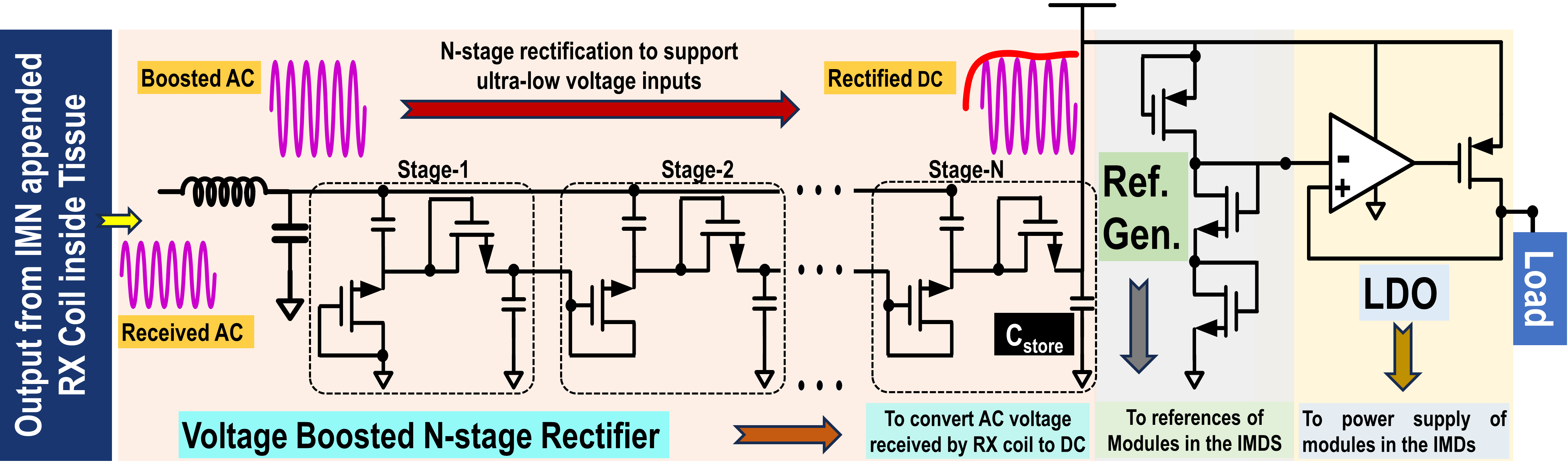}
\caption{\asif{Overview of N-stage Energy Harvester: This harvester system comprises (a) an N-stage CMOS rectifier with a voltage booster to convert the AC signal from the RX coil into a rectified DC output, (b) a reference generator to provide a stable reference voltage for other modules within the system, and (c) low dropout regulator (LDO) to regulate the rectified DC voltage to a constant, usable level.}
}
\label{all}
\end{figure*}

With the TX coil in air and the RX coil embedded in biological tissue, the characteristics of the TRX coils are significantly influenced by the properties of the medium. Biological tissues are inherently conductive and frequency-dependent media, which induces eddy currents in the inductive link, affecting the magnetic field distribution and energy dissipation in the form of heat. Consequently, the corresponding power loss can be described using Farday’s law and cole-cole expression (detailed explanation in Appendix D) \cite{pozar2011microwave,gabriel1996dielectric}.

\begin{equation}
P_{\text{loss, tissue}}  \propto \sigma \omega^2
\label{eq}
\end{equation}

where, $\sigma$ denotes the frequency-dependent conductivity of the media and $\omega$ is the operating angular frequency. 
The proportionality constant depends on the transmitter geometry, the coil separation, and the media properties. In general, an optimal tissue-channel model accounts for the frequency-dependent dielectric behavior of all different layers of the tissue environment. Thus, the equivalent distributed complex impedances originating from the tissue environment in Fig.~\ref{trmed} will be crucial for optimal NRIC links. But it is difficult to take into account these effects in theoretical approximations.
As per literature, a distributed RC model in \cite{cho2007human}, FEM model in \cite{xu2010electric}, transmission line models in \cite{callejon2011study}, and heterogeneous tissue model in \cite{mao2018five} are suitable for use in circuit simulations. This study emphasizes heterogeneous tissue-induced effects by dividing them into a mesh of complex transverse and longitudinal impedances depicted as $Z_{Vn}$ and $Z_{Hn}$. The conductivity and relative permittivity of the relevant layer can be referenced from \cite{gabriel1996dielectric} and the combined tissue-induced losses can be conveniently obtained from CAD tools like ANSYS HFSS in terms of modified ABCD parameters of TRX coils as shown in Fig.~\ref{trmed} as well as in (27). 
 \begin{equation}
[T_{TRX}^{'}] =\begin{bmatrix}
A_{coil}^{'} &B_{coil}^{'}  \\
C_{coil}^{'} &D_{coil}^{'}  \\
\end{bmatrix}\label{eq}
\end{equation}
 
\asif{Thus, the effects of the biological tissue environment can be incorporated into the modified transmission parameters by replacing $A_{coil}$, $B_{coil}$, $C_{coil}$, $D_{coil}$ by $A_{coil}^{'}$, $B_{coil}^{'}$, $C_{coil}^{'}$, $D_{coil}^{'}$ respectively in (17)-(20). These modified parameters are then converted back to S-parameters for the TRX coils (shown in details in Appendix B), which ensures that the design specifications are effectively met.} This approach further optimizes the modeling of required coil geometry while accounting for the influence of the tissue medium. Additionally, it provides updated input and output impedances ($Z_{in1}$ and $Z_{in2}$) for the TRX coils, which are critical for the accurate design of the IMN component.

\subsection{Analysis of L-type IMN}
To understand the paramount impact of IMN in TRX coil design, an example is provided in Fig.~\ref{LIMN_0}. Having $L_{1}$ = $L_{2}$ = 400 nH, $R_{1}$ = $R_{2}$ = $0.5~\Omega$ and $k$ = 0.1 for a pair of coils to operate at 20 MHz with both TX and RX ports terminated at $50~\Omega$, the conventional approach can be adding a SS compensation network with $C_{TX}$ = $C_{RX}$ = 158.3 pF as shown in scenario-1 of Fig.~\ref{LIMN_0}(a). But PTE is not maximum for this configuration as evident from scenario-2 of Fig.~\ref{LIMN_0}(a) after combining a IMN with $L_{s1}$ = $L_{s2}$ = 166.7 nH and $C_{p1}$ = $C_{p2}$ = 293.6 pF. Now instead of adding resonant capacitors followed by IMN, we can directly append the IMN validated in the same test case as shown in Fig.~\ref{LIMN_0}(b). The initial step is to ensure $L_{opt}$ = $L_{1}$ = $L_{2}$ = 400nH which gives the best PTE at 20 MHz. Then the next and final step is to apply the L-type IMN where $C_{s1}$ = $C_{s2}$ = 52.7 pF and $C_{p1}$ = $C_{p2}$ = 109.1 pF to minimize  $S_{11,link}$ and $S_{22,link}$ which will further enhance the PTE. Here, $S_{11,link}$ and $S_{22,link}$ denote input return loss (RL) and output RL respectively for the entire NRIC link. Ultimately, the same PTE level is reached for both the cases with a much simplified approach in the latter method as shown in Fig.~\ref{LIMN_0}(b).

Now, established IMN topologies encompass L, pi, and T types and out of these, L-type IMN is the simplest to implement using only capacitors or only inductors or a combination of both. Four possible L-type configuration has been summarized in Fig.~\ref{LIMN_1} where [$T_{link}$] represents the transmission matrix for the entire NRIC link and $Z_{s_{j}}=j\omega L_{s_{j}}$ or $\frac{1}{j\omega C_{s_{j}}}$, $Z_{p_{j}}=j\omega L_{p_{j}}$ or $\frac{1}{j\omega C_{p_{j}}}$. Depending on the choice of inductors or capacitors, each configuration gives rise to to $2^{4}=16$ possibilities to accomplish the IMN part. Thus, for all the cases, $4 \times 16 = 64$ distinct L-type IMN networks are available for appropriate deployment and [$T_{link}$] combining the IMNs is presented in (32).
\begin{equation}
[T_{link}]=\begin{bmatrix}
A_{link} & B_{link} \\ 
C_{link} & D_{link}
\end{bmatrix}\label{eq}
\end{equation}

Furthermore, setting $S_{11,link}$ = $S_{22,link}$ = 0 and transforming [$T_{link}$] presented in (32) to the S-matrix \cite{10182236,omi2021novel}, we can determine the IMN elements ($L_{s_{j}}$, $L_{p_{j}}$,$C_{s_{j}}$, $C_{p_{j}}$) solving (33) and (34) simultaneously.
\begin{equation}
Z_{p2}A_{link} + B_{link} = Z_{p1}Z_{p2}C_{link} + Z_{p1}D_{link}
\label{eq}
\end{equation}
\begin{equation}
Z_{p1}D_{link} + B_{link} = Z_{p1}Z_{p2}C_{link} + Z_{p2}A_{link}\label{eq}
\end{equation}

Thus, the analytical design process is deemed complete paving the way for maximal PTE via the NRIC link.

\subsection{N-stage Energy Harvester System}
Upon confirming the highest possible PTE, the output of the RX coil connects to an Energy Harvester system, which plays a pivotal role in powering various modules within the implanted device. The harvesting system must operate efficiently under low power conditions and within the constraints of the WPT system. N-stage CMOS rectifier is the key compoenent of this harvesting system \cite{Guler2017,mohan2022design,lyu2020synchronized}. Several design considerations are critical for the rectifier to ensure the reliable performance of the overall system \cite{Martins2021, Karami2019}.


Threshold voltage, reverse leakage, parasitics, and input impedance variations mostly impact the rectifier's power conversion capability. Within such constraints, numerous studies have been reported in the literature to enhance the performance of rectifiers by applying various topologies. Among these notable techniques, adaptive threshold compensation \cite{ Hameed2015}, auxiliary circuit technique \cite{ 10549952}, body-bias \cite{Moghaddam2017}, active technique \cite{ Yu2022, Karimi2024, karimi2024wireless}, self $V_{th}$ compensation \cite{ Saffari2019}, the half-wave voltage doubler \cite{ Jiang2017}, full-wave voltage doubler \cite{ 10479226}, cross-coupled \cite{Xu2021, Xu2022}, Dickson multiplier with diode-connected transistors \cite{oh201232dbm}, self-biased \cite{XuRectifier2022} and reconfigurable architectures \cite{Abouzied2017, Zeng2019} stand out as prominent examples. Considering the ultra-low power applications enabled by the subthreshold region operation in \cite{oh201232dbm}, we have adopted this topology to facilitate the energy harvesting system. Another compelling rationale for selecting this topology is its suitability for addressing the limitations on the power received at the RX coil of the NRIC link, such as safety concerns like tissue heating, restrictions on implant size, and the influence of the surrounding conductive tissue medium. Even under conditions of ultra-low wireless power is received at the RX coil, this approach will ensure the availability of required DC voltage at the rectifier output.

Now, with $N$, representing the number of stages, $V_{RX}$ as the amplitude of the received signal at the RX coil, $V_{T}$ as the thermal voltage, $I_{0}$ as the zero-th order modified Bessel function of the first kind, $Z_{in,EH}$ as the input impedance of the energy harvesting system, $R_{rect}$ as the effective input resistance of the rectifier and $C_{rect}$ as the effective input capacitance of the rectifier including the parasitic capacitance of the transistors and storage capacitors, we can deduce the values of the rectified DC voltage, $V_{out}$ as shown in \asif{(32)} and $R_{rect}$ as well $C_{rect}$ in \asif{(33) and (34)} respectively.

\begin{equation}
V_{out}=2NV_{T}ln\left ( \frac{I_{0}V_{RX}}{V_{T}} \right )\label{eq}
\end{equation}
\begin{equation}
R_{rect}=\frac{\left| Z_{in,EH}\right|^{2}}{R_{in,EH}}\label{eq}
\end{equation}
\begin{equation}
C_{rect}=\frac{-X_{in,EH}}{2\pi f\left| Z_{in,EH}\right|^{2}}\label{eq}
\end{equation}

where, $R_{in,EH}=Re\left\{ Z_{in,EH}\right\}$ and $X_{in,EH}=Im\left\{ Z_{in,EH}\right\}$.

Based on these design equations, the available DC voltage can be estimated, which is cardinal for powering various modules in the IMDs.

\section{The Systematic Design Methodology}
\subsection{BWPT with TRX Coils}

\begin{figure}[t]
\centerline{\includegraphics[width=\columnwidth]{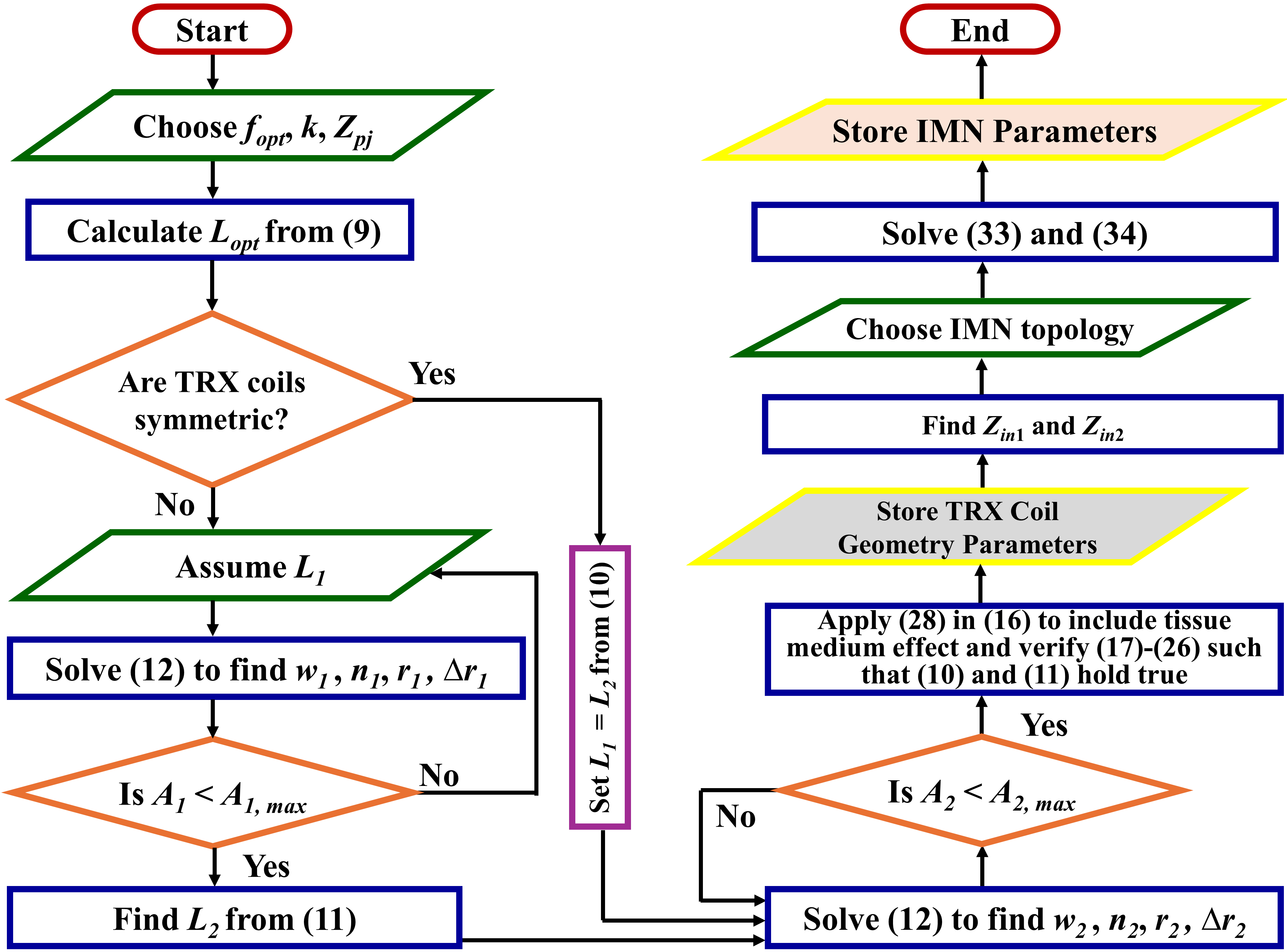}}
\caption{Step-by-step analytical methodology for designing TRX coil pairs of varying sizes (both symmetric and asymmetric). For implantable systems, the RX coil is typically smaller than the TX coil, adding complexity to the design process while aiming to achieve the maximum PTE.}
\label{flowchart1}
\end{figure}
\begin{figure}[t]
\centerline{\includegraphics[width=1\columnwidth]{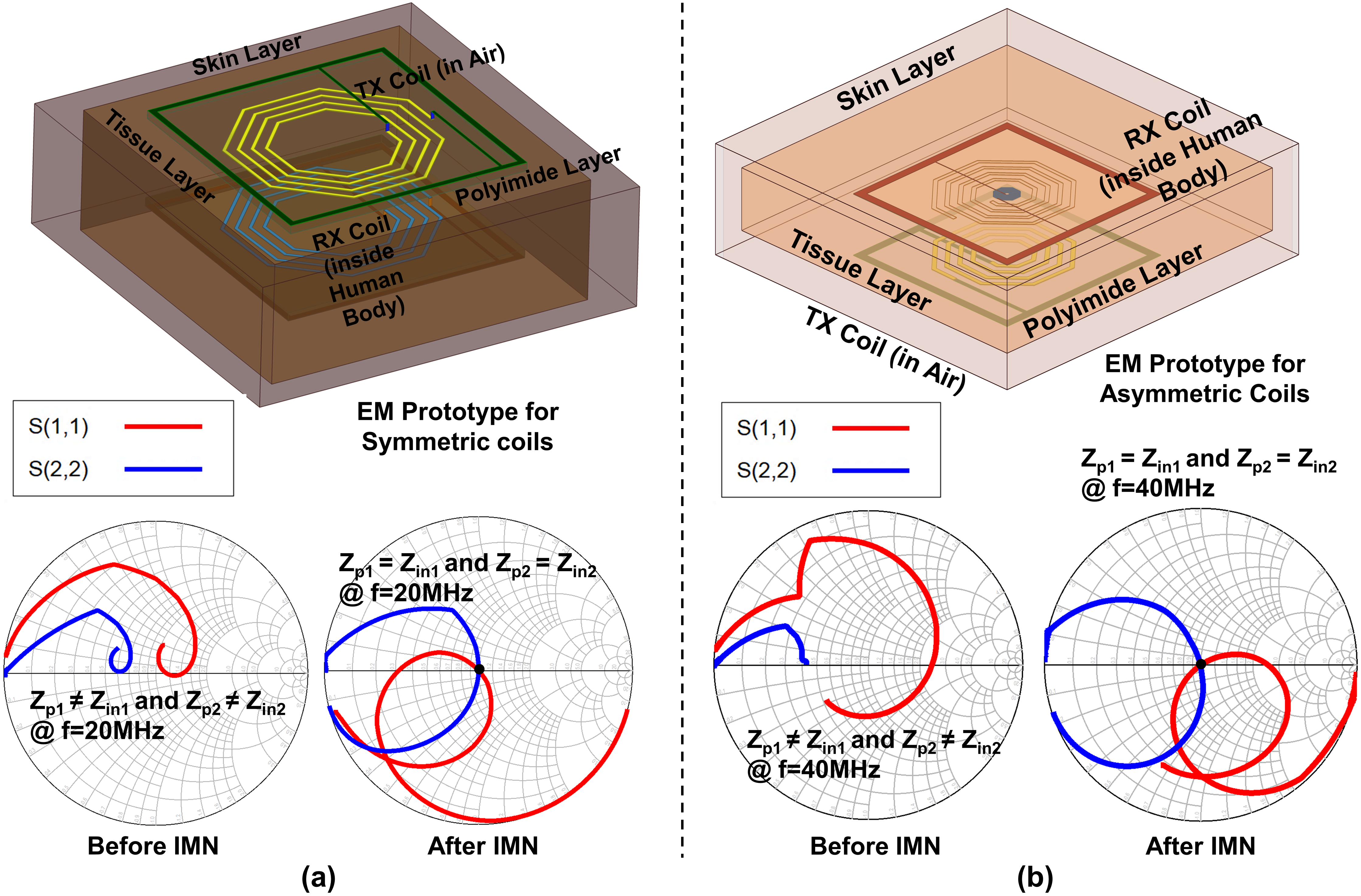}}
\caption{Design examples of TRX coils with $k$ = 0.2: (a) symmetric types at 20 MHz and (b) asymmetric types at 40 MHz. EM models are obtained for both the types where $L_{opt}$ = $L_{1}$ = $L_{2}$ = 400.4 nH in (a) and $L_{1}$ = 525.7 nH, $L_{2}$ = 80 nH in (b). With the suitable L-type IMNs (case-2 for (a) and case-1 for (b) from Fig.~\ref{LIMN_1}), the Smith chart signifies the required matching.}
\label{sym}
\end{figure}
\begin{figure}[t]
\centerline{\includegraphics[width=1\columnwidth]{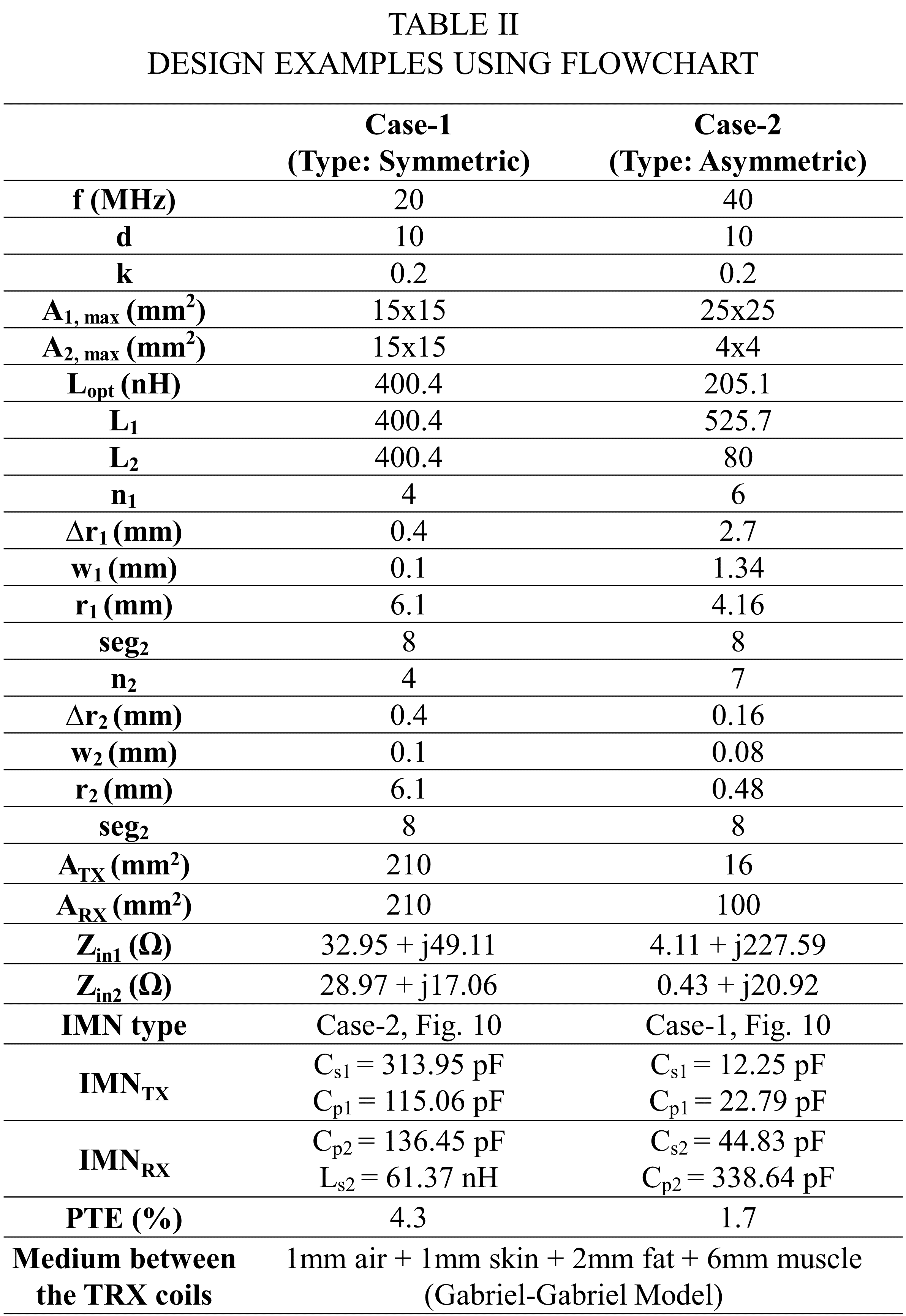}}
\label{tab2}
\end{figure}


Now, in accordance with the previously developed sections in II (A-E), Fig.~\ref{flowchart1} depicts the flowchart showing the proposed step-by-step design formulation incorporating the effect of tissue medium. To demonstrate the proposed concept, one symmetric two-coil BWPT system, and one asymmetric type system are designed to achieve maximal PTE within the constraints of human body operation. The design details are mentioned in Table II which follows the proposed flowchart in Fig.~\ref{flowchart1}. Based on these values in Table II, the coil layout is shown in Fig.~\ref{sym}. Simulation results from the Smith chart, as also showcased in Fig.~\ref{sym}, indicate the lossy nature of the conductive tissue environment through the observed impedance values at 20MHz. The calculated components for the IMNs allow the system to maintain the desired transmission characteristics as also seen from the Smith chart in Fig.~\ref{sym} for both the cases. 

Therefore, this proposed design model will be highly beneficial for any designer wanting to maximize the BWPT capability of NRIC links.

\subsection{Design Considerations for Harvesting Circuits}
\begin{figure}[t]
\centerline{\includegraphics[width=1\columnwidth]{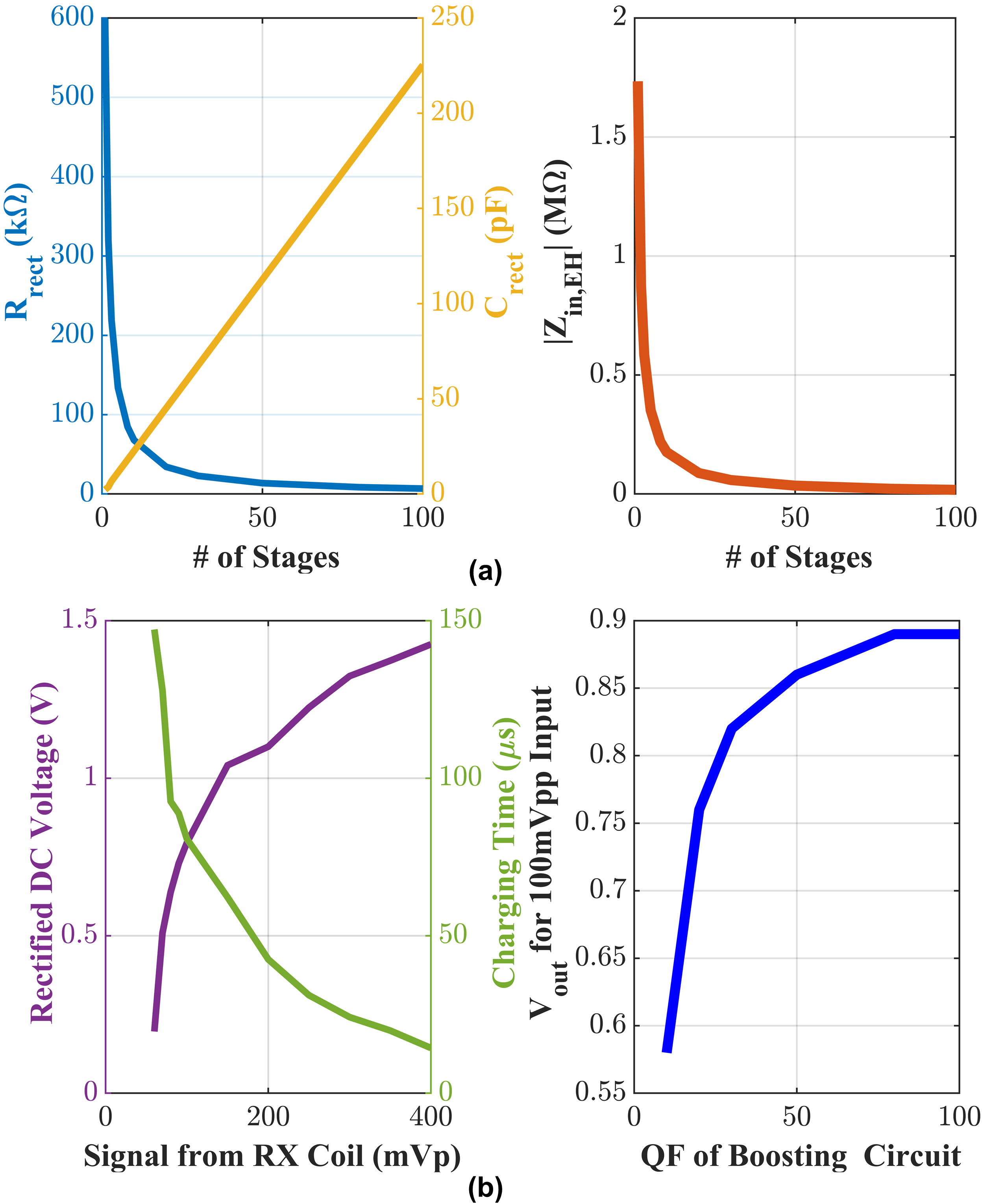}}
\caption{Design Space exploration for rectifier with the boosting circuit: (a) input impedance variation as a function of the number of stages, (b) rectified DC voltage and corresponding charging time, and (c) output voltage as a function of the quality factor of the boosting circuit. It is essential to consider these factors simultaneously to enable ultra-low voltage operation.}
\label{ehe}
\vspace{-3mm}
\end{figure}
\begin{figure}[htbp]
\centerline{\includegraphics[width=1\columnwidth]{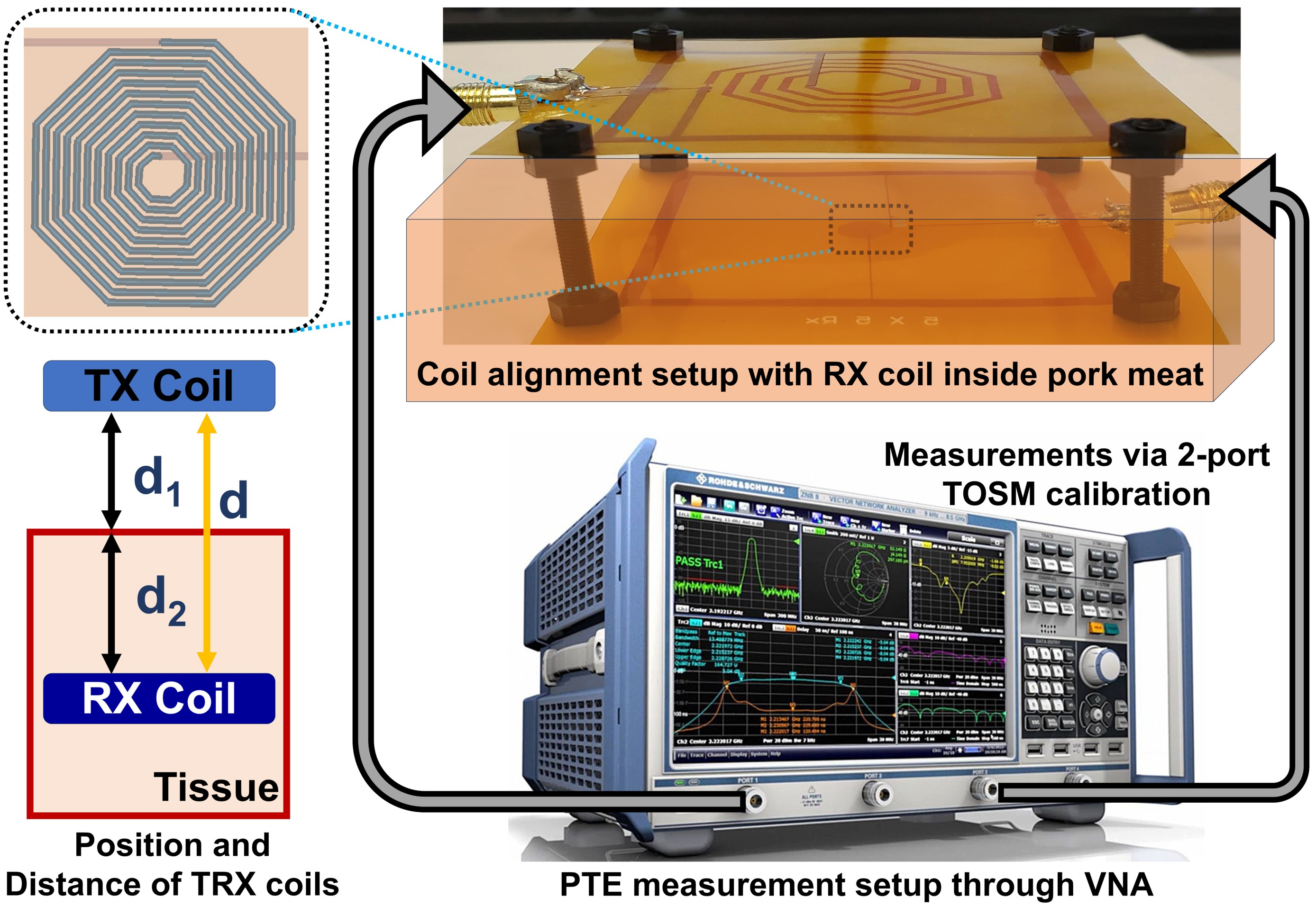}}
\caption{Fabricated prototyes (analytically determined using the proposed flowchart) with alignment and positioning configurations to set up the TRX coils properly for estimating the link performance with 2-port TOSM calibration.}
\label{meas1}
\end{figure}
The circuit model based on sub-threshold operation \cite{oh201232dbm} doesn't provide an in-depth analysis or guideline on how to obtain the optimum number of stages, how to choose the Q (Q-factor) of the boosting circuit, and how the input impedance of the rectifier plays a critical role in multi-stage rectification. So, it is important to develop a guideline to incorporate these factors when designing an N-stage rectifier. Additionally, the capacitors are replaced and approximated with the corresponding transistor configuration to make the system miniaturized, which would be highly conducive to implants. 

From the previous sub-section in III(A), it is conspicuous that the appropriate set of TRX coil geometries dictates the whole NRIC link design process. Based on the available $V_RX$, the proposed approach to designing an N-stage energy harvesting system is as follows.

\begin{enumerate}
    \item Choose the supply voltage at the TX side.
    \item Initiate the NRIC link design process to get the desired $V_{RX}$ based on section III (A).
    \item Assume Q to realize an LC resonating boosting circuit.
    \item Assume width, $W$ and length $L_{T}$ for each transistor to ensure operation in the sub-threshold region
    \item Assume $N$ and find $Z_{in,EH}$. 
    \item Determine $N$ such that $R_{rect}$ and $C_{rect}$ shown in \asif{(33)} and \asif{(34)} complex conjugately match with the tissue impedance.
    \item Find output impedance of the energy harvesting system and approximate charging time from it.
    \item Check whether the charging time is practically realizable with the chosen $N$. If yes, continue. If not, change the Q and/or the transistor sizing and repeat from \asif{step 3}.
    \item Calculate $V_{out}$ from \asif{(32)}.
    \item Check whether the DC voltage is sufficient at the designed frequency. If not, change the supply voltage at the TX side and repeat from (1). If yes, store the values of all $W$ and $L_{T}$.
\end{enumerate}

Utilizing the above-mentioned design methodology, a design space is explored at 100 MHz as an example in Fig.~\ref{ehe}. This same approach can investigate the design space across any desired frequency range, as defined by the NRIC link parameters. For optimal performance facilitating low-power applications, it is observed that higher $N$, higher $R_{rect}$, lower $C_{rect}$, lower charging time, and higher $Q$ are required. Thus, the optimum number of $N$ can be determined by simultaneously considering the input and output impedance of the rectifier, maintaining the sub-threshold region of operation. 

\begin{figure}[t]
\centerline{\includegraphics[width=1\columnwidth]{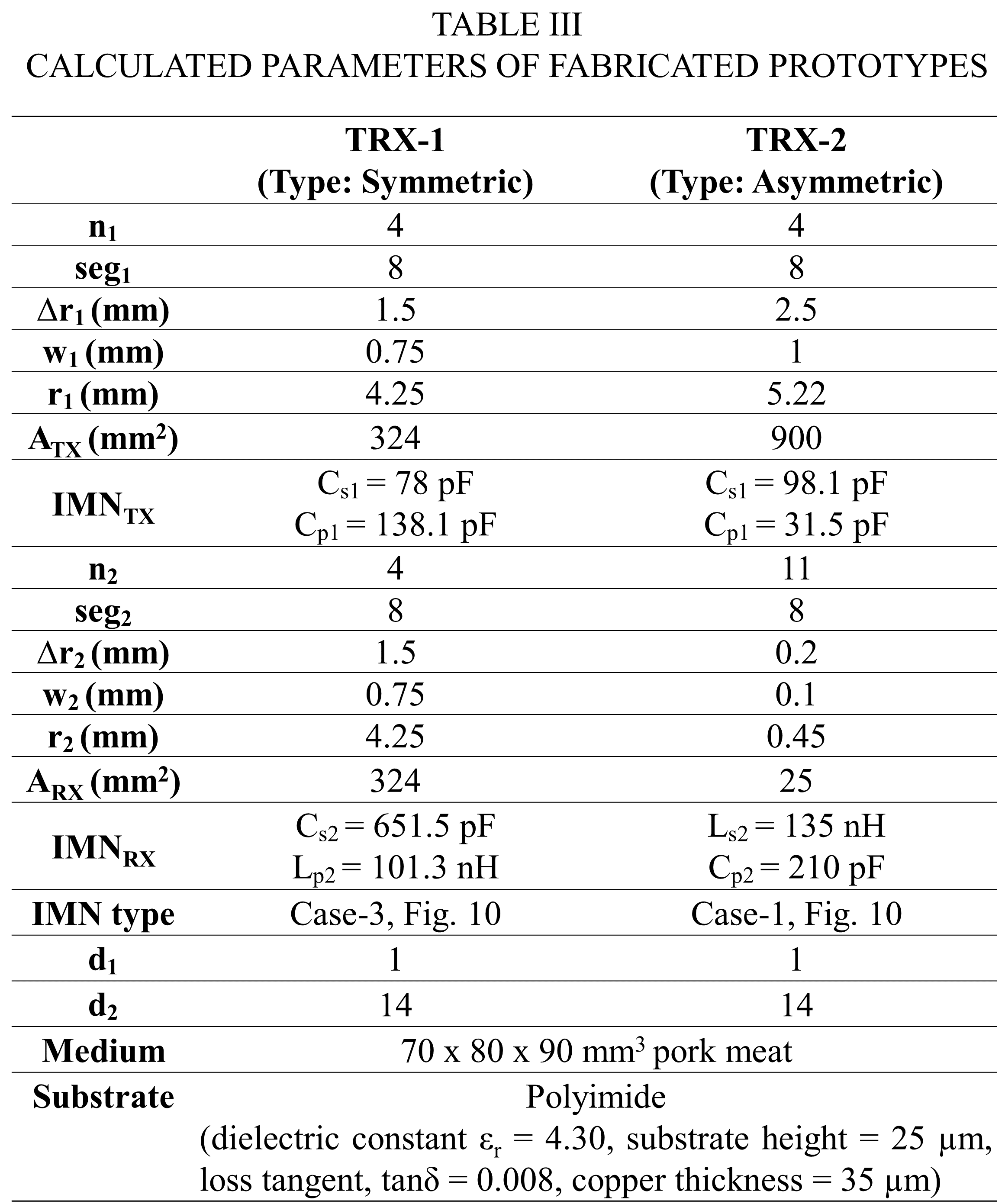}}
\label{meas1}
\end{figure}

\begin{figure}[t]
\centerline{\includegraphics[width=1\columnwidth]{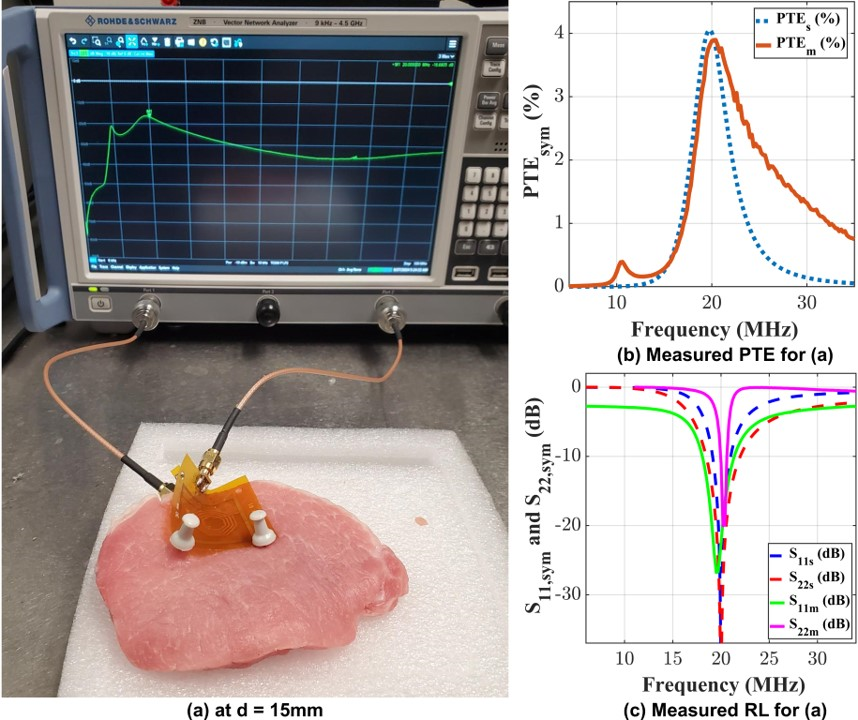}}
\vspace{-1mm}
\caption{Performance of type-1 (symmetric) prototype: (a) demonstration of $S_{21}$ measurement, (b) PTE comparison, and (c) corresponding matching performance. To assess simulated vs measured, the subscript `s’ indicates simulated result, and `m’ indicates measured data in (b)-(c) for (a). 
}
\label{meas3}
\end{figure}
\begin{figure}[t]
\centerline{\includegraphics[width=1\columnwidth]{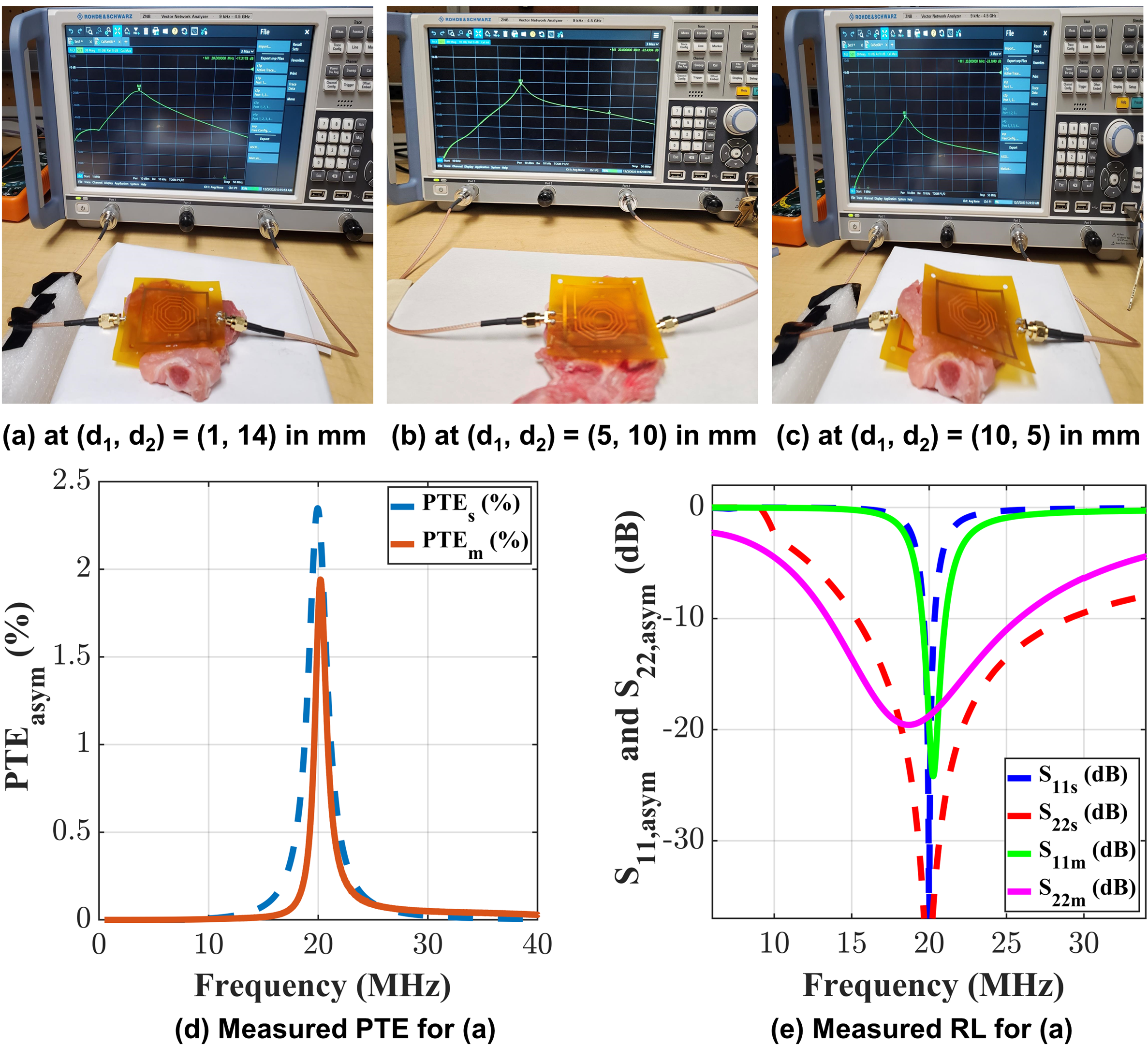}}
\vspace{-1mm}
\caption{Performance of type-2 (asymmetric) prototype: (a)-(c) demonstration of $S_{21}$ measurements, (d) PTE comparison, and (e) corresponding matching performance. 
Here, the subscript `s’ indicates simulated result, and `m’ indicates measured data in (d)-(e) for (a). 
}
\label{meas2}
\vspace{-5mm}
\end{figure}

\begin{figure*}[t]
\centerline{\includegraphics[width=1\textwidth]{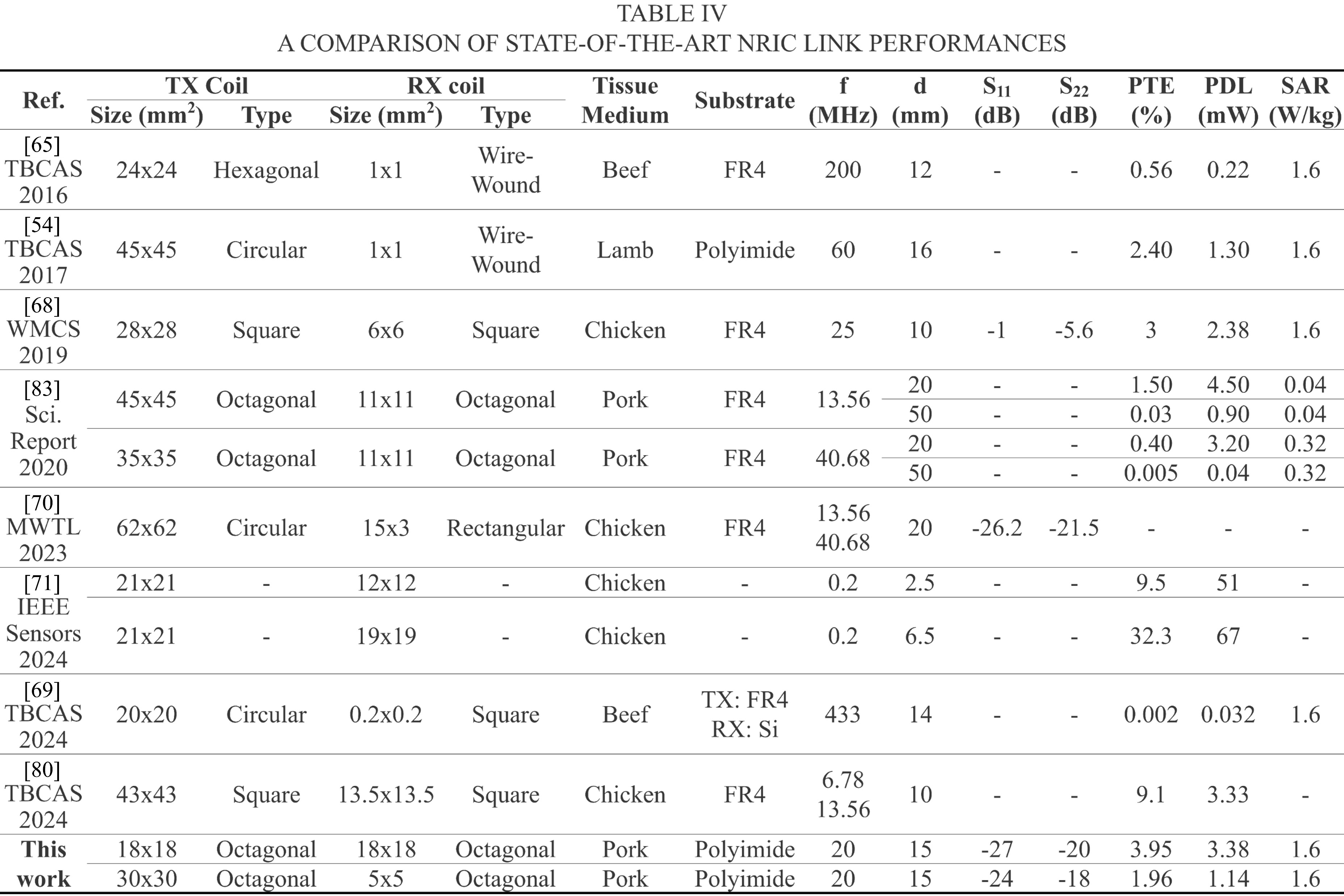}}
\label{comp1}
\end{figure*}

\section{The Measurement Results and Discussion}
\subsection{Performance of NRIC Link}
To validate the proposed concept, two NRIC links (type-1: symmetric TRX coils and type-2: asymmetric TRX coils) including the IMNs are designed with a target to achieve maximal PTE where pork meat serves as tissue medium. The operational frequency is chosen as 20MHz for its favorable tissue penetration characteristics and compliance with IEEE 802.15.6 standards. Polyimide is used as the optimal flexible substrate for both type-1 and type-2 TRX coils, highlighting its advantageous properties, including high dielectric strength, thermal stability, and biocompatibility—crucial for efficient BWPT \cite{zhang2018wireless}. Electromagnetic simulations are performed using Ansys HFSS, with biological tissue modeled by the Gabriel-Gabriel Model \cite{gabriel1996dielectric}, prior to finalizing the coil PCBs. \asif{Also, $70 \times 80 \times 90 mm^{3}$ cube fresh pork meat was used as an ex-vivo model due to its close resemblance to human tissues in dielectric properties (relative permittivity and conductivity), ensuring comparable results. Studies with implantable antennas have already confirmed similar responses between porcine and human tissues across the frequency range of 300 MHz–3 GHz \cite{karacolak2012dielectric}, making pork meat a practical and effective surrogate for experimental validation. 
}

\subsubsection{Symmetric TRX Coils}
According to Fig.\ref{flowchart1}, the optimal inductance is determined to be $L_{opt} = 400.4$ nH for $k = 0.1$ and a RX coil area $A_{RX1}$ smaller than the maximum allowable area ($A_{RX1(max)} = 18 \times 18$ $mm^{2}$). Among several potential solutions to achieve the highest possible PTE within the specified area constraint, the coil dimensions are detailed in Table III. The corresponding measurement setup for aligning the fabricated PCBs is illustrated in Fig.~\ref{meas1}. Furthermore, an LC-combined L-type network is validated by solving equations (30) and (31). Measurements, conducted using 2-port TOSM calibration with a vector network analyzer (Rohde \& Schwarz ZNB 4PORT VNA), are presented in Fig.\ref{meas3}(a), depicting the BWPT scenarios. As shown in Fig.\ref{meas3}(b)-(c), the fabricated prototypes exhibit excellent transmission and reflection characteristics, achieving approximately 4\% PTE with a miniaturized $18 \times 18 \times 0.025$ $mm^{3}$ RX coil size (including substrate thickness). The required impedance matching is confirmed by input/output RL values $>$ 20 dB (where RL $>$ 10 dB is ideal). Additionally, the measured results closely align with simulated outcomes at 20 MHz, further validating the proposed design methodology for symmetrical coil-based NRIC links.

\subsubsection{Asymmetric TRX Coils}
Based on Fig.~\ref{flowchart1}, the optimum inductance is found as $L_{opt}$ = 400.4 nH for $k$ = 0.1 and $A_{RX2}$ $<$ $A_{RX2(max)}$ (= $5 \times 5$ $mm^{2}$). Out of multiple solutions to ensure the best possible PTE option within the area constraint, the coil dimension details are again elaborated in Table III. Further, an LC-combined L-type network is validated while solving (30) and (31). Again, the measurements performed with 2-port TOSM calibration using a vector network analyzer (Rohde \& Schwarz ZNB 4PORT VNA) are showcased in Fig.~\ref{meas2}(a)-(c), indicating different scenarios of BWPT. It is apparent from Fig.~\ref{meas2}(d)-(e) that the fabricated prototypes have remarkable transmission as well as reflection parameters evident by $\approx$ 2\% PTE considering a miniaturized $5 \times 5 \times 0.025$ $mm^{3}$ RX coil size (including substrate thickness). In this regard, the required matching is also ensured, which can be inferred from the fact that RL is $>$ 15 dB in both the TRXs (ideally, RL $>$ 10 dB is imperative). Moreover, the measured outputs are compared with the simulated results at 20 MHz which shows good agreement. Consequently, the experimental research confirms the capability of the proposed design methodology for asymmetrical coils-based NRIC link. 

\subsubsection{SAR Constrained PDL}
\begin{figure}[t]
\centerline{\includegraphics[width=1\columnwidth]{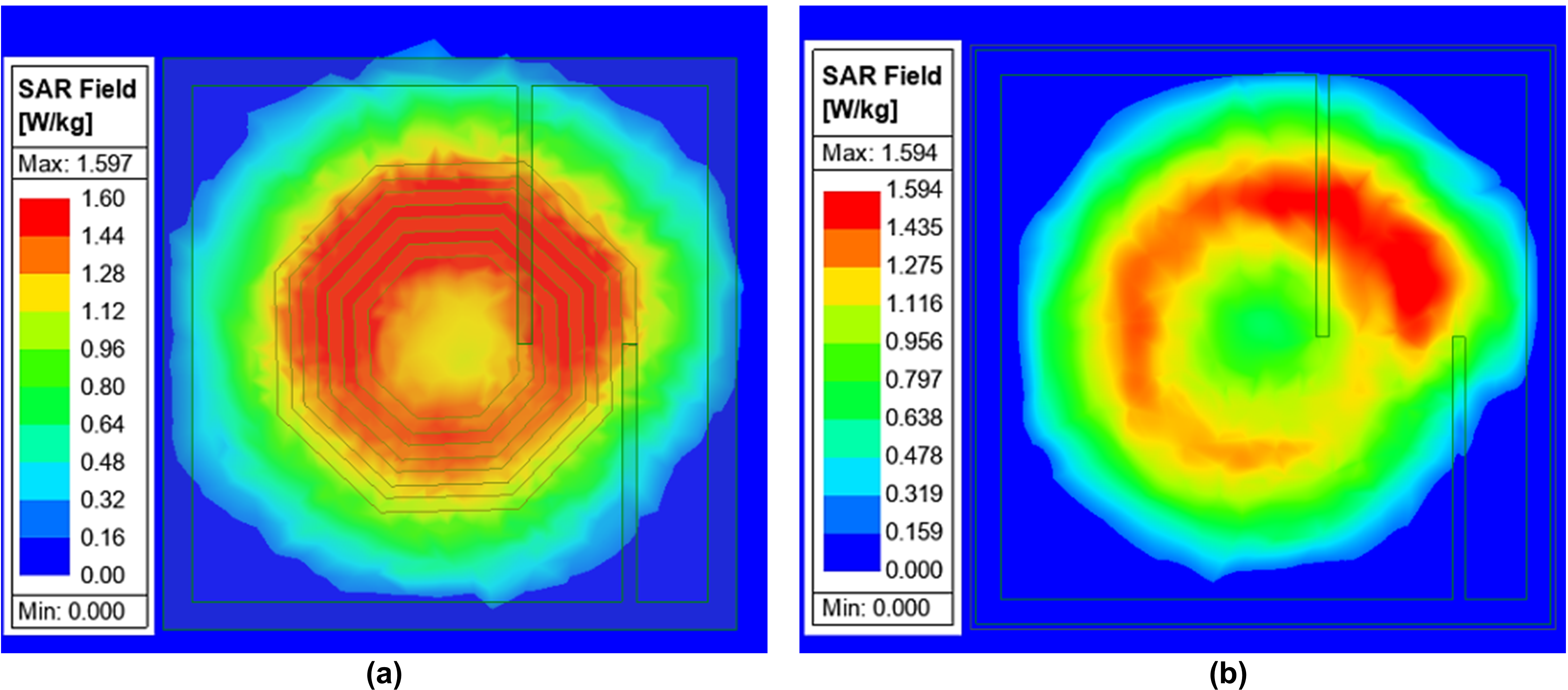}}
\caption{SAR estimate for: (a) type-1 TRX coils and (b) type-2 TRX coils. The transmitted powers from TX coils to reach this limit of 1.6 W/kg are found to be 84.6 mW and 57 mW respectively for these TRX coils. Thus, the maximum received power at the RX coils is SAR constrained.}
\label{sar}
\vspace{-3mm}
\end{figure}
Ensuring safety during BWPT is paramount, as the maximum power transmitted by an NRIC link is constrained by the maximum averaged Specific Absorption Rate (SAR) specified by safety standards \cite{1626482, 1266063, ICNIRP1998}. SAR quantifies the rate at which electromagnetic energy is absorbed by the body, with basic restrictions designed to prevent thermal effects caused by RF energy exposure. For frequencies between 100 kHz and 6 GHz, SAR limits take precedence over reference levels for field strength and power density and must not be exceeded \cite{1626482}. According to IEEE guidelines, SAR limits are set at 1.6 W/kg for any 1 g of tissue and 4 W/kg for any 10 g of tissue. The ICNIRP guidelines \cite{ICNIRP1998} specify SAR limits of 4 W/kg for any 10 g of tissue in the hands, wrists, feet, and ankles, and 2 W/kg for any 10 g of other tissues.

Adhering to these safety frameworks, we evaluated SAR field distribution across heterogeneous tissue layers (2 mm skin, 2 mm fat, and 10 mm muscle) at 20 MHz to assess electromagnetic field exposure, as shown in Fig.\ref{sar}. With 84.6 mW transmitted by type-1 TX coils, the peak simulated average SAR observed along the tissue surface was 1.597 W/kg, precisely matching the maximum safety threshold, as depicted in Fig.\ref{sar}(a). Similarly, Fig.~\ref{sar}(b) shows that the maximum allowable transmitted power for type-2 TX coils is 57 mW, achieving a similar SAR limit of 1.594 W/kg. Given the Power Transfer Efficiency (PTE) of 4\% for type-1 TX coils and 2\% for type-2 TX coils, the maximum PDL inside the pork meat can be considered 3.38 mW for symmetric TRX coils and 1.14 mW for asymmetric TRX coils. Thus, SAR limits combined with PTE determine the optimal PDL.

Finally, to further emphasize the effectiveness of the TRX coil prototypes, the manuscript includes a detailed comparison in Table IV, benchmarking the measured performance of the fabricated TRX coils against state-of-the-art designs reported in the literature. The data clearly demonstrates that the proposed NRIC link design meets or exceeds the requirements across various comparable parameters, providing a vivid assessment of their performances.

\subsection{Performance of Energy Harvester IC}
\begin{figure}[t]
\centerline{\includegraphics[width=1\columnwidth]{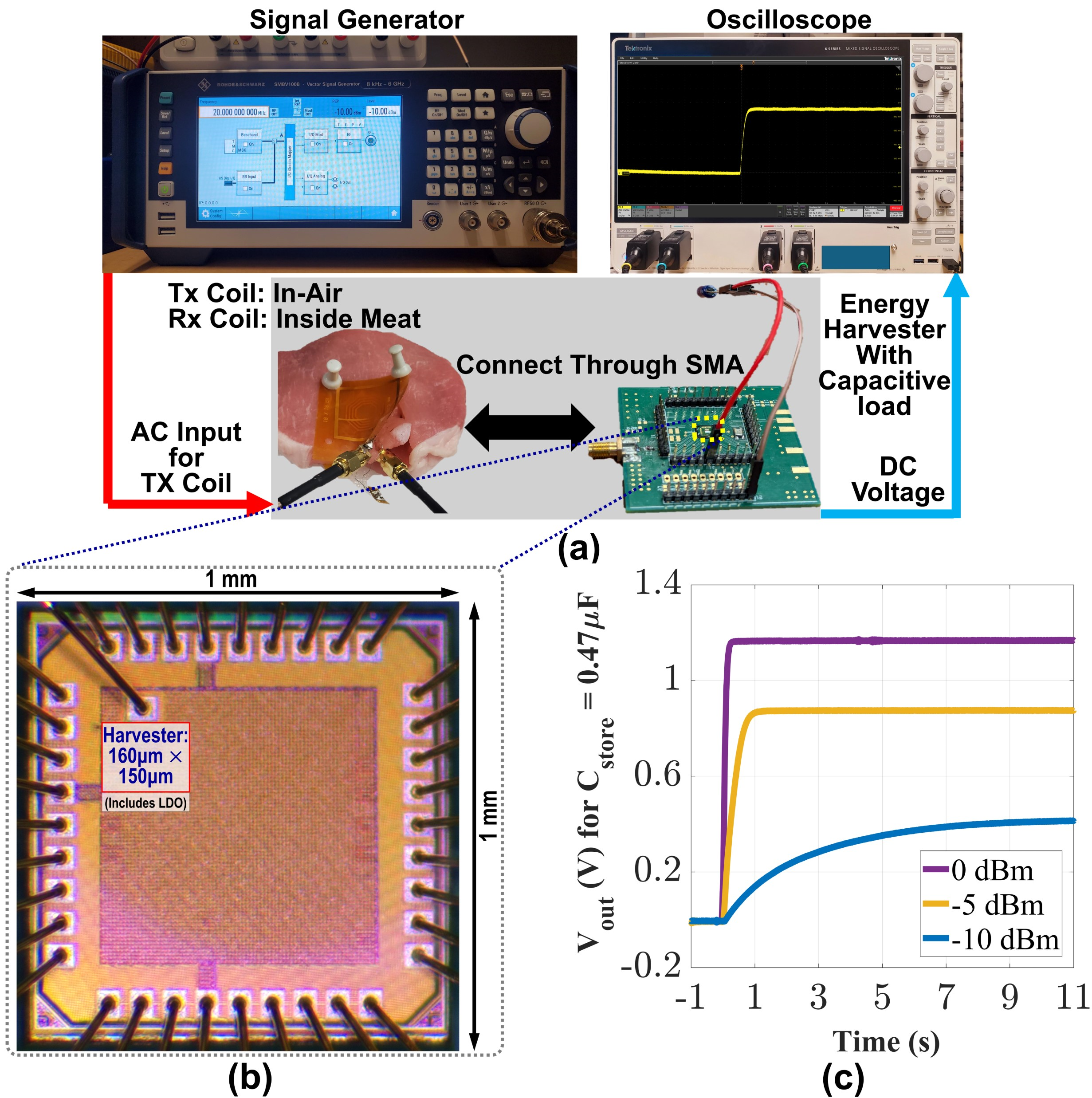}}
\caption{Measurement setup for the energy harvesting system with a 30-stage rectifier on a test-chip: (a) voltage rectification demonstration, (b) chip micrograph of the harvester unit, and (c) time-domain waveforms of $V_{out}$ at various power levels. The setup illustrates the integrated performance of the harvester chip in conjunction with TRX coil PCBs.}
\label{eh1}
\end{figure}

\begin{figure}[t]
\centerline{\includegraphics[width=0.8\columnwidth]{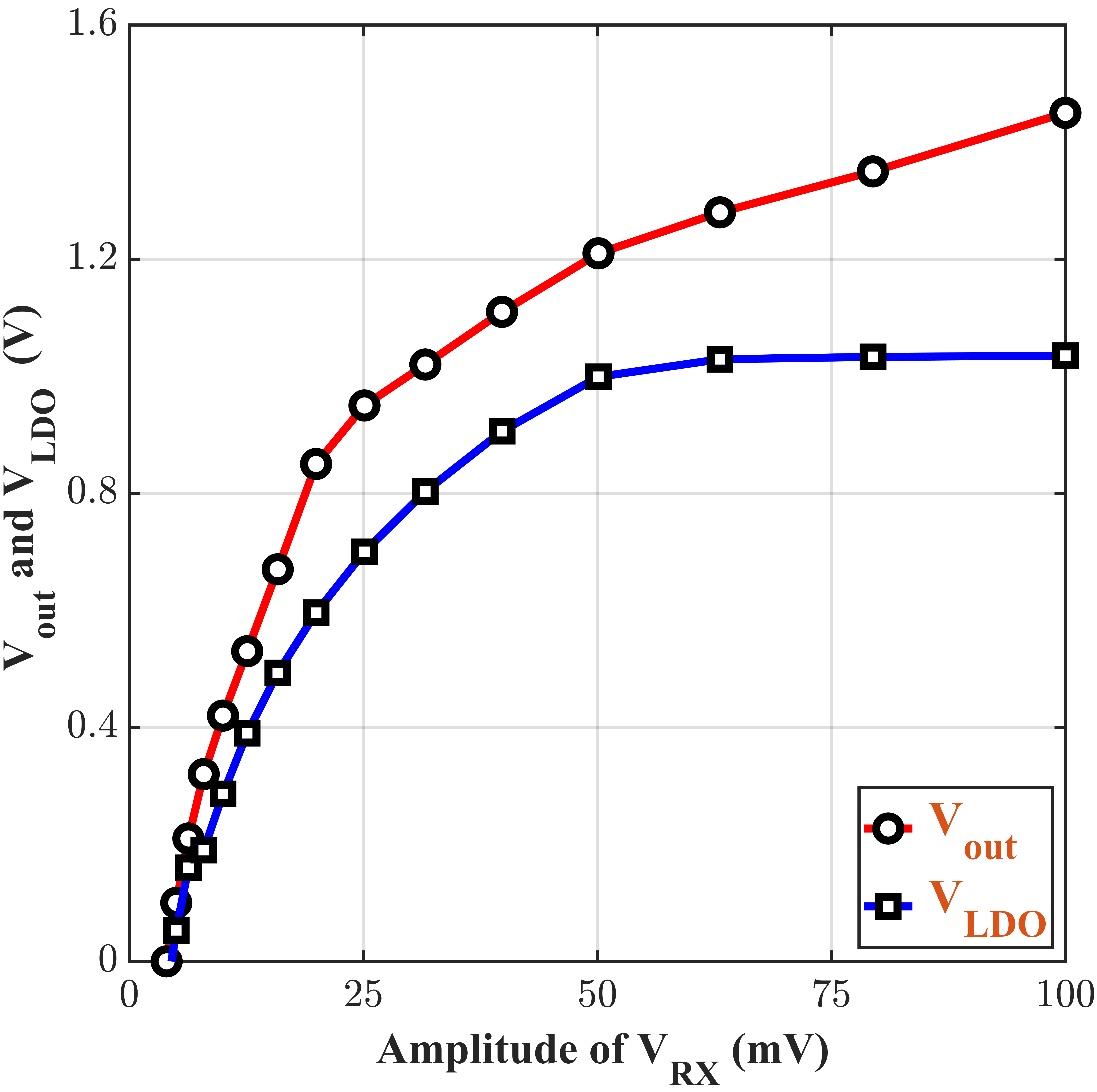}}
\vspace{-3mm}
\caption{Rectifier and LDO output as a function of varying $V_{RX}$ levels. The LDO output stabilizes at approximately 1.05 V when the $V_{RX}$ amplitude exceeds 60 mV for an external storage capacitor of 0.47 $\mu $F.}
\label{eh2}
\vspace{-5mm}
\end{figure}

The measurement setup for the energy harvester system, depicted in Fig.~\ref{eh1}(a), showcases the integration of the energy harvester system with the NRIC link. The energy harvester IC was fabricated using the TSMC 65nm CMOS process, as evidenced by the die micrograph in Fig.~\ref{eh1}(b). 
The chip's core area, which includes a 30-stage rectifier, reference generator, and low dropout regulator (LDO), covers 0.024 $mm^{2}$ (160 $\mu m$ $\times$ 150 $\mu m$), while the overall chip occupies 1 $mm^{2}$. For the 65nm rectifier design, n-MOS transistors with dimensions of 200nm/60nm were utilized, while the rectifier capacitors were replaced and approximated using transistors configured with p-MOS and n-MOS sizes of 2 $\mu m$ / 2 $\mu m$. The boosting circuit incorporates an LC network consisting of an off-chip 6.33 $ \mu H$ inductor and an on-chip 10 pF metal-insulator-metal (MIM) capacitor. Under these conditions, an optimal stage count of N = 30 achieves a DC output exceeding 1 V from an initial 50 mV input at the RX coils.

Among the two types of NRIC links, the asymmetric TRX coils were selected to interface with the chip. This choice aligns with practical scenarios where the TX coil typically has a larger area, while the RX coil is smaller. Additionally, under SAR-constrained conditions, the power delivered in this configuration is $<$ 0.54 dBm, setting the range for safe operation. Now, in this configuration, the signal generator is directly connected to the TX coils of the NRIC link, facilitating BWPT through pork meat, as evident from the time-domain waveforms in Fig.~\ref{eh1}(c). Also, as shown in Fig.~\ref{eh1}(c), the output DC voltage is strongly influenced by the input power level. For example, when the input power reaches 0 dBm at the design frequency, with an external capacitive load of 0.47 $\mu F$, the energy harvester system delivers a DC output voltage of 1.2 V. Further analysis of the system’s performance indicates that increasing the input power levels boosts the output voltage and accelerates the charging time. Additionally, the received voltage level through the NRIC link also impacts the rectifier and the LDO output. As seen from Fig.~\ref{eh2}, when the magnitude of $V_{RX}$ surpasses 60 mV, the LDO output stabilizes at approximately 1.05 V. 

Hence, the overall results underscore the efficacy of the designed energy harvester, particularly when interfaced with the TRX coil PCBs, ensuring reliable operation under varying input conditions and emphasize the system’s potential to be a critical component in future biomedical and low-power applications.

\section{Conclusion}
This paper introduces a new systematic design paradigm that effectively maximizes BWPT in area-constrained two-coil NRIC links. 
Several examples have been provided to demonstrate the effectiveness of the proposed technique. One of the main challenges of this study was obtaining high PTE despite coil alignment within the complex multi-layered tissue model, which necessitated minimal electromagnetic (EM) optimization in the post-design phase. The NRIC links successfully powered the type-1 and type-2 RX coils, delivering a maximum of 3.38 mW and  1.14 mW, respectively, at the 20 MHz optimal frequency without surpassing the SAR limits. Overall, the implemented proof-of-concept prototypes \asif{substantiate} excellent agreement between EM simulated and measured responses.
\asif{Additionally, the 30-stage energy harvester system enabled by the 20 MHz NRIC link delivered a stable 1V output considering trade-offs among the number of stages, input impedance, and charging time.} 
As future work, multiple-coil NRIC links with automatically reconfigurable IMNs} at various TX-RX alignments will be explored, and the current work will be expanded to integrate multiple levels of LDO outputs for 
supporting multiple power domains in an SoC.

\appendices
\section{Derivation of $f_{opt}$}
This section describes the derivation of $f_{opt}$ shown in Fig.~\ref{intro2}(c) and discussed in Section II (A).  The two-coil link for the TRX coils can be transformed into a T network with mutual inductance \(M\), and primary/secondary inductances of \(L_1 - M\) and \(L_2 - M\) respectively which gives rise to the characteristic Z-matrix shown in (1). Fig.~\ref{ztrx} depicts the two-port network for the TRX coils, defined by its S-matrix and connected to port impedances $Z_{p1}$ and $Z_{p2}$. Now, from the conversion between S and Z-parameters \cite{pozar2011microwave},

\begin{equation}
S_{11,TRX} = \frac{(Z_{11,coil} - Z_{p1})(Z_{22,coil} + Z_{p2}) - Z_{12,coil}Z_{21,coil}}{\Delta_Z}
\label{eq}
\end{equation}
\begin{equation}
S_{12,TRX} = S_{21,TRX} = \frac{2\sqrt{Z_{p1}Z_{p2}} Z_{21,coil}}{\Delta_Z}
\label{eq}
\end{equation}
\begin{equation}
S_{22,TRX} = \frac{(Z_{11,coil} + Z_{p1})(Z_{22,coil} - Z_{p2}) - Z_{12,coil}Z_{21,coil}}{\Delta_Z}
\label{eq}
\end{equation}

where,
\[
\Delta_Z = (Z_{11,coil} + Z_{p1})(Z_{22,coil} + Z_{p2}) - Z_{12,coil}Z_{21,coil}
\]

Also, from the (1) in Section II (A),

\begin{equation}
Z_{11,coil} = R_{1} + j \omega L_{1}
\label{eq}
\end{equation}
\begin{equation}
Z_{12,coil} = Z_{21,coil} = j \omega M
\label{eq}
\end{equation}
\begin{equation}
Z_{22,coil} = R_{2} + j \omega L_{2}
\label{eq}
\end{equation}

Now using (38)-(40) in (36), a simple algebraic manipulation produces (2). In the absence of IMNs, the TRX coils exhibit poor matching. As a result, the frequency at which $S_{21,TRX}$ reaches its maximum can be considered the optimal operating frequency for the NRIC link.Further, to maximize $S_{21,TRX}$ at $f_{opt}$,

\begin{equation}
\frac{\mathrm{d} }{\mathrm{d} f}(\left | S_{21,TRX} \right |)=0
\label{eq}
\end{equation}

Solving (41) leads to the derivation of$f_{opt}$ as follows.

\begin{equation}
f_{\text{opt}} = \frac{1}{2\pi} \sqrt{\frac{(R_1 + Z_{p1})(R_2 + Z_{p2})}{L_1 L_2 - M^2}}
\label{eq}
\end{equation}

which is the same as shown in (7). 

\section{Validation of TRX Coil Parameters}

This section illustrates the validation and derivation of circuit parameters of the TRX coils in terms of S-parameters after determining the coil geometry as discussed in Section II (C). From the Z to ABCD parameter conversion in (16),

\begin{equation}
A_{\text{coil}} = \frac{R_1 + j\omega L_1}{j\omega M}
\label{eq}
\end{equation}
\begin{equation}
C_{\text{coil}} = \frac{1}{j\omega M}
\label{eq}
\end{equation}
\begin{equation}
D_{\text{coil}} = \frac{R_2 + j\omega L_2}{j\omega M}
\label{eq}
\end{equation}

Now from (43)-(45) after some algebraic manipulation leads to determination of $L_{1}$, $L_{2}$, $R_{1}$, $R_{2}$, and $k$ as deduced in (46)-(50).

\begin{equation}
L_{1}=\frac{Im\left\{ \frac{A_{coil}}{C_{coil}}\right\}}{2\pi f}\label{eq}
\end{equation}
\begin{equation}
R_{1}=Re\left\{ \frac{A_{coil}}{C_{coil}}\right\}\label{eq}
\end{equation}
\begin{equation}
L_{2}=\frac{Im\left\{ \frac{D_{coil}}{C_{coil}}\right\}}{2\pi f}\label{eq}
\end{equation}
\begin{equation}
R_{2}=Re\left\{ \frac{D_{coil}}{C_{coil}}\right\}\label{eq} 
\end{equation}
\begin{equation}
k=\left [ Re\left\{ A_{coil}\right\}Re\left\{ D_{coil}\right\} \right ]^{-\frac{1}{2}}\label{eq} 
\end{equation}

Again, these transmission parameters of the TRX coils can be transformed into corresponding S-parameters for unequal $Z_{pj}$ as shown in (51)-(53).

\begin{equation}
A_{coil}=\frac{\sqrt{Z_{p1}}(pv+x)}{2\sqrt{Z_{p2}}S_{12,TRX}}\label{eq}
\end{equation}
\begin{equation}
C_{coil}=\frac{(qv-x)}{2\sqrt{Z_{p1}Z_{p2}}S_{12,TRX}}\label{eq}
\end{equation}
\begin{equation}
D_{coil}=\frac{\sqrt{Z_{p2}}(qu+x)}{2\sqrt{Z_{p1}}S_{12,TRX}}\label{eq}
\end{equation}

Further, equating $A_{coil}, C_{coil}$ and $ D_{coil}$ from (51)-(53) into (46)-(50), validates (17)-(21).

\section{PTE Formula for the NRIC Link}
\begin{figure}[t]
\centerline{\includegraphics[width=0.6\columnwidth]{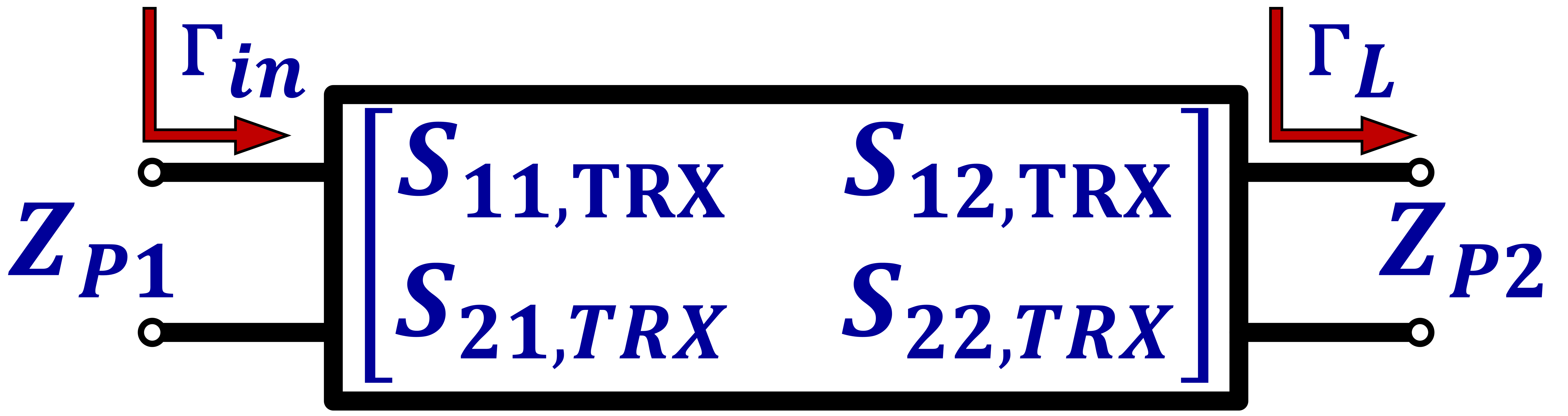}}
\caption{Two-port network for the TRX coils showing the S-parameters}
\label{ztrx}
\vspace{-5mm}
\end{figure}

This section formulates the relation between PTE and S-parameters discussed several times in Section II. If $\Gamma_{in}$ and $\Gamma_{L}$ shown in Fig.~\ref{ztrx} represent the reflection coefficients seen looking at the input of the two-port TRX coils and looking toward the load respectively, then the PTE of the TRX coils, $PTE_{TRX}$ can be defined as \cite{pozar2011microwave},

\begin{equation}
PTE_{TRX} = \frac{|S_{21,TRX}|^2 (1 - |\Gamma_L|^2)}{(1 - |\Gamma_{\text{in}}|^2)|1 - S_{22,TRX} \Gamma_L|^2}
\label{eq}
\end{equation}

where,
\[
\Gamma_{\text{in}} = S_{11,TRX} + \frac{S_{21,TRX}^2 \Gamma_L}{1 - S_{22,TRX}\Gamma_L}
\]

Assuming ideal complex conjugate matching at TX and RX side, the maximum achievable PTE for the NRIC link can be derived as follows \cite{zargham2014fully,pozar2011microwave}. 

\begin{equation}
PTE_{max,link} = K_{r} - \sqrt{K_{r}^2 - 1}
\label{eq}
\end{equation}

where \(K_{r}\) is given by:

\begin{equation}
K_{r} = \frac{1 + a + b + c}{2|S_{21,TRX}^2|}
\end{equation}
\begin{equation}
a = |S_{11,TRX}S_{22,TRX} - S_{21,TRX}^2|^2
\end{equation}
\begin{equation}
b = - |S_{11,TRX}|^2
\end{equation}
\begin{equation}
c = - |S_{22,TRX}|^2
\end{equation}

After appending the IMN as discussed in Section II (E), the two-port S-parameters for the TRX coils in (55)-(59) can be replaced by $S_{11,link}$, $S_{22,link}$ and $S_{21,link}$, which are the corresponding S-parameters of the NRIC link (TRX coils + IMN). Also, for such loss-less matching, $S_{11,link} = S_{22,link} = 0$. This leads to the derivation of the the IMN parameters from (29) and (30). Also, the commonly used PTE formula for the NRIC link is obtained as follows.

\begin{equation}
PTE_{link} = |S_{21,link}|^2 \leq PTE_{max,link}
\label{eq}
\end{equation}

The equality in (60) holds true if the NRIC link is designed at $f = f_{opt}$ and IMN is implemented as ideally as possible. Now, (60) is valid for $Z_{p1} = Z_{p2}$ ($= $50 $\Omega $ typically). 
But when $Z_{p1} \neq Z_{p2} \neq$50 $\Omega $, we can utilize the method in \cite{mirbozorgi2018power} and the modified PTE formula for the NRIC link is given in (61).

\begin{equation}
PTE_{link} = \gamma \times|S_{21,link}|^2
\label{eq}
\end{equation}

where,

case-1: $\gamma = \frac{Z_{p2}}{Z_{p1}} $ for $Z_{p2} > 50, Z_{p1} < 50$

case-2: $\gamma = \left(\frac{Z_{p2}}{50}\right) \times \left(\frac{Z_{p1}}{50}\right) $ for $Z_{p2} > 50, Z_{p1} > 50$

case-3: $\gamma = \frac{Z_{p1}}{Z_{p2}} $ for $Z_{p2} < 50, Z_{p1} > 50$

case-4: $\gamma = \left(\frac{50}{Z_{p2}}\right) \times \left(\frac{50}{Z_{p1}}\right) $ for $Z_{p2} < 50, Z_{p1} < 50$

\section{Power Loss in Biological Tissue Medium}

This section provides an estimate of power loss in the biological tissue medium as discussed briefly in Section II (D). According to Faraday’s law, a time-varying magnetic field induces an electric field which is the source of power dissipation inside the conductive tissue medium. Subsequently, from Maxwell's equation \cite{barbruni2024frequency}, 

\begin{equation}
\nabla \times \mathbf{E} = -\frac{\partial \mathbf{B}}{\partial t} \propto \omega I
\label{eq}
\end{equation}

where, $E$ indicates electric field, $B$ is the magnetic field, $\omega $ denotes the operating angular frequency and $I$ is the flowing current. Thus it is clear from (58) that,

\begin{equation}
|E| \propto \omega
\label{eq}
\end{equation}

Further, the frequency-dependent loss inside the conductive tissue medium can be described by the following relation \cite{poon2010optimal}.

\begin{equation}
P_{\text{loss, tissue}} = \frac{\omega}{2} \int_V \text{Im}\{\epsilon\} |E|^2 \, dV
\label{eq}
\end{equation}

where, $Im\left\{ \epsilon\right\}$ represents the imaginary component of the permittivity of the medium, which varies with the frequency-dependent conductivity, $\sigma$ and complex permittivity, $\epsilon$ of the tissue. The Cole-Cole model for biological tissues provides a widely accepted approximation for permittivity, incorporating both conductive and dielectric properties \cite{gabriel1996dielectric} as follows.

\begin{equation}
\epsilon = \epsilon_{\infty} + \frac{\epsilon_s - \epsilon_{\infty}}{1 + (j \omega \tau)^{1-\alpha}} + \frac{\sigma}{j \omega \epsilon_0}
\label{eq}
\end{equation}

The imaginary component of the permittivity can be effectively approximated by the final term under the condition $\omega \tau <<$ 1, a criterion that is valid for all biological tissues within the specified frequency range of interest. Notably, $\omega \tau <$ 0.05 for frequencies up to 500 MHz across all relevant tissue types \cite{gabriel1996dielectric}.
Therefore, simultaneously applying (59) and (61) in (60), we can estimate the power loss in tissue conductive region as highlighted in (27).

\bibliographystyle{IEEEtran}
\bibliography{bare_jrnl}

\end{document}